\documentclass[12pt, DIV10,a4paper]{article}

\usepackage{color}
\usepackage{float}
\usepackage[bf,small]{caption2}
\usepackage{comment}
\usepackage{amsmath}
\usepackage{bigints}
\usepackage{rotating, booktabs}
\usepackage{amsopn}
\usepackage{amsfonts}
\usepackage{amsthm}
\usepackage{amssymb}
\usepackage{amsbsy}
\usepackage{gensymb}
\usepackage{multirow}
\usepackage{titling}
\usepackage{natbib}
\usepackage{tocbibind}
\usepackage[a4paper,colorlinks,breaklinks,bookmarksopen,bookmarksnumbered]{hyperref}
\usepackage{epsfig,psfrag}
\usepackage{graphicx}
\usepackage{amsmath}
\usepackage{epstopdf}
\usepackage{tabularx}
\usepackage{subfigure}
\usepackage[mathscr]{euscript}
\usepackage[margin=1in]{geometry}
\usepackage{xr-hyper}

\usepackage{amsfonts} 
\usepackage{calc}
\usepackage{titlesec}

\newcommand{\ueq}[1][]{%
  \if\relax\detokenize{#1}\relax
    \sbox0{$\underbrace{=}_{}$}%
    \mathrel{\mathmakebox[\wd0]{=}}
  \else
    \mathrel{\underbrace{=}_{\mathclap{#1}}}
  \fi}

\newcommand{\bzero}{\boldsymbol{0}}

\newcommand {\ctn}{\cite}

\usepackage{url}
\renewcommand{\t}{\ensuremath{\theta}}

\newcommand{\btheta}{\boldsymbol{\theta}}

\newcommand{\btau}{\boldsymbol{\tau}}

\newcommand{\bgamma}{\boldsymbol{\gamma}}
\newcommand{\bSigma}{\boldsymbol{\Sigma}}

\newcommand{\bpi}{\boldsymbol{\pi}}

\newcommand{\bmu}{\boldsymbol{\mu}}
\newcommand{\bnu}{\boldsymbol{\nu}}
\newcommand{\bzeta}{\boldsymbol{\zeta}}

\newcommand{\bomega}{\boldsymbol{\omega}}

\newcommand{\bB}{\boldsymbol{B}}

\newcommand{\bA}{\boldsymbol{A}}
\newcommand{\bS}{\boldsymbol{S}}

\newcommand{\by}{\boldsymbol{y}}

\newtheorem{theorem}{Theorem}

\newtheorem{definition}[theorem]{Definition}

\renewcommand{\t}{\ensuremath{\theta}}

\newcommand{\topline}{\hrule height 1pt width \textwidth \vspace*{2pt}}
\newcommand{\botline}{\vspace*{2pt}\hrule height 1pt width \textwidth \vspace*{4pt}}
\newtheorem{algo}{Algorithm} 

\begin{document}

\title{\vspace{-0.8in}
\textbf{IID Sampling from Intractable Multimodal and Variable-Dimensional Distributions}}
\author{Sourabh Bhattacharya\thanks{
Sourabh Bhattacharya is an Associate Professor in Interdisciplinary Statistical Research Unit, Indian Statistical
Institute, 203, B. T. Road, Kolkata 700108.
Corresponding e-mail: sourabh@isical.ac.in.}}
\date{\vspace{-0.5in}}
\maketitle%
	
\begin{abstract}
\ctn{Bhatta21} has introduced a novel methodology for generating $iid$ realizations from any target distribution on the Euclidean space,
irrespective of dimensionality. In this article, our purpose is two-fold. We first extend the method for obtaining $iid$ realizations from general multimodal distributions,
and illustrate with a mixture of two $50$-dimensional normal distributions. Then we extend the $iid$ sampling method for fixed-dimensional distributions 
to variable-dimensional situations and illustrate with a variable-dimensional normal mixture modeling of the well-known ``acidity data", with further
demonstration of the applicability of the $iid$ sampling method developed for multimodal distributions.
\\[2mm]

\end{abstract}
	

\section{Introduction}
\label{sec:introduction}
Statistical problems dealing with unknown dimensionality is now ubiquitous in the literature. Various examples include mixtures with unknown number of components,
change point analysis problems with unknown number and locations of change points, autoregression in time series with unknown order of autoregression, covariate selection 
problems in parametric and nonparametric regression, factor analysis with unknown dimension of the latent factor matrix, spatial point processes with unknown locations
and number of points, nonparametric regression with unknown number of basis functions, and so on. We refer to \ctn{Das19} for a more comprehensive overview, along with
relevant references. The classical statistical paradigm ignores uncertainty about the unknown dimension by fixing its value by means of some existing (usually ad-hoc) 
model selection criterion. It is the Bayesian paradigm that is better equipped to deal with unknown dimension by proposing a prior for the same and then formulating the joint
posterior distribution, to proceed with inference.

However, traditional, fixed-dimensional Markov Chain Monte Carlo (MCMC) applications are clearly not applicable to variable-dimensional problems. In this regard,
\ctn{Green95} introduced reversible jump Markov Chain Monte Carlo (RJMCMC) that converges in principle to the desired variable-dimensional (posterior) distribution.
Unfortunately, such is the inefficiency of RJMCMC in practice that by now researchers have almost completely shunned the method, and have taken recourse to
fixing the unknown dimension in the same vein as in classical statistics. 

An appropriate and indeed a far more efficient and powerful methodology for handing variable-dimensional distributions has been introduced by \ctn{Das19},
who generalize the fixed dimensional transformation based Markov Chain Monte Carlo (TMCMC) of \ctn{Dutta14} to the variable-dimensional setup.
The new method has been referred to as transdimensional transformation based Markov Chain Monte Carlo (TTMCMC). The key idea is to update most or all
the unknowns using suitable deterministic transformations of low dimensional random variables, leading to drastic
effective dimension reduction operated by fixed dimensional moves. For numerous advantages of TTMCMC over RJMCMC, see \ctn{Das19}.
So far TTMCMC has been very successfully applied to mixtures with unknown number of components (\ctn{Das19}), variable selection in parametric and nonparametric
setups including the ``large $p$ small $n$" paradigm (\ctn{Minerva21}) and nonparametric spatio-temporal contexts (\ctn{Das20} and \ctn{Bhattacharya21}, the latter
consisting of about $400$ unknown dimensions).

Now, since all MCMC algorithms are asymptotic, ascertainment of convergence is a serious issue and there exists a plethora of empirical and ad-hoc 
methods for convergence diagnosis, none of which is satisfactory enough. It is thus certainly worth investigating if the convergence issue can be
eradicated altogether. In this regard, the idea of perfect sampling, introduced by \ctn{Prop96}, is a step forward in the right direction.
Although hitherto regarded as only a proof of concept and not meant for serious business, \ctn{Bhatta21} has been able to create a 
novel method for $iid$ sampling, showing, with ample illustrations, 
that the idea can be judiciously exploited to generate $iid$ samples of any desired size, from any distribution on the Euclidean space, irrespective of dimension.

In this article, we explore the prospects of $iid$ sampling from variable-dimensional distributions by adopting and extending the method proposed by
\ctn{Bhatta21}. For illustration, we choose the acidity data that \ctn{Richardson97}, \ctn{Bhattacharya08} 
and \ctn{Das19} model by normal mixtures with unknown number of components. Since multimodality of the posteriors of the mixture model parameters is a well-known
phenomenon, this requires us to first extend the theory and method of \ctn{Bhatta21} to generically accommodate multimodality. We develop the extension and
provide illustration with a $50$-dimensional, two-component normal mixture. Integrating the ideas with the key concepts of \ctn{Bhatta21} results in a methodology
that is capable of generating $iid$ realizations from generic multimodal and variable-dimensional distributions, which we illustrate with the acidity data.

The rest of our article is organized as follows. We begin by an overview of the $iid$ sampling idea of \ctn{Bhatta21} in in Section \ref{sec:idea}.
In Section \ref{sec:multimodal} we develop the theory, method and the algorithm for generating $iid$ samples from general multimodal distributions
on Euclidean spaces with arbitrary dimensions. The important issue of obtaining the modes of the multimodal target distributions, necessary for $iid$ sampling, 
is taken up in Section \ref{sec:modes}.
A simulation study involving $iid$ sample generation from a two-component, $50$-dimensional normal mixture, is detailed in Section \ref{sec:simstudy_multmodal},
to illustrate our theory and method.
In Section \ref{sec:vardim}, we further extend our $iid$ sampling theory and method to simulate from variable-dimensional target distributions, providing the general algorithm
for the purpose, and in Section \ref{sec:vardim_example}, illustrate our methodology with a variable-dimensional normal mixture model for the well-recognised acidity data.
Finally, we summarize our ideas and make concluding remarks in Section \ref{sec:conclusion}.

\section{An overview of the $iid$ sampling idea}
\label{sec:idea}

For $\btheta=(\theta_1,\ldots,\theta_d)^T\in\mathbb R^d$, let $\pi(\btheta)$ be the target distribution from which $iid$ realizations are required. 
Note that the distribution can be represented as
\begin{equation}
	\pi(\btheta)=\sum_{i=1}^{\infty}\pi(\bA_i)\pi_i(\btheta),
	\label{eq:p1}
\end{equation}
where $\bA_i$ are disjoint compact subsets of $\mathbb R^d$ such that $\cup_{i=1}^{\infty}\bA_i=\mathbb R^d$, and
\begin{equation}
	\pi_i(\btheta)=\frac{\pi(\btheta)}{\pi(\bA_i)}I_{\bA_i}(\btheta), 
	\label{eq:p2}
\end{equation}
is the distribution of $\btheta$ restricted on $\bA_i$; $I_{\bA_i}$ being the indicator function of $\bA_i$.
In (\ref{eq:p1}), $\pi(\bA_i)=\int_{\bA_i}\pi(d\btheta)\geq 0$. Clearly, $\sum_{i=1}^{\infty}\pi(\bA_i)=1$.

The key idea of generating $iid$ realizations from $\pi(\btheta)$ is to randomly select $\pi_i$ with probability $\pi(\bA_i)$ and then to perfectly simulate
from $\pi_i$. 

\subsection{Choice of the sets $\bA_i$}
\label{subsec:choice_sets}
For some appropriate $d$-dimensional vector $\bmu$ and $d\times d$ positive definite scale matrix $\bSigma$,
we shall set $\bA_i=\{\btheta:c_{i-1}\leq (\btheta-\bmu)^T\bSigma^{-1}(\btheta-\bmu)\leq c_i\}$ for $i=1,2,\ldots$, where $0=c_0<c_1<c_2<\cdots$. 
Note that $\bA_1=\{\btheta:(\btheta-\bmu)^T\bSigma^{-1}(\btheta-\bmu)\leq c_1\}$, and for $i\geq 2$, 
$\bA_i=\{\btheta:(\btheta-\bmu)^T\bSigma^{-1}(\btheta-\bmu)\leq c_i\}\setminus\cup_{j=1}^{i-1}\bA_j$.
Observe that ideally one should set 
$\bA_i=\{\btheta:c_{i-1}<(\btheta-\bmu)^T\bSigma^{-1}(\btheta-\bmu)\leq c_i\}$, but since
$\btheta$ has continuous distribution in our setup, we shall continue with $\bA_i=\{\btheta:c_{i-1}\leq (\btheta-\bmu)^T\bSigma^{-1}(\btheta-\bmu)\leq c_i\}$.

Thus, the first member of the sequence of sets $\bA_i$; $i\geq 1$, is a closed ellipsoid, while the others are closed annuli, the regions between two 
successive concentric closed ellipsoids.
The compact ellipsoid $\bA_1$ tends to support the modal region of the target distribution $\pi$ and for increasing $i\geq 2$, the compact 
annuli $\bA_i$ tend to support the tail regions
of the target distribution. The radii $\sqrt{c_i}$; $i\geq 1$, play important role in the efficiency of the underlying perfect simulation procedure, and hence 
must be chosen with care.

The choices of $\bmu$ and $\bSigma$ will be based on TMCMC estimates of the mean (if it exists, or co-ordinate-wise median otherwise) 
and covariance structure of $\pi$ (if it exists, or some appropriate scale matrix otherwise).

\subsection{Selecting $\pi_i$ and perfect sampling from $\pi_i$}
\label{subsec:ptmcmc}
Recall that for perfect sampling from $\pi(\btheta)$ we first need to select $\pi_i$ with probability proportional to $\pi(\bA_i)$ 
for some $i\geq 1$, and then need to sample from $\pi_i$ in the perfect sense.
As shown in \ctn{Bhatta21}, up to a normalizing constant, $\pi(\bA_i)$ can be approximated arbitrarily accurately by Monte Carlo averaging of samples
drawn uniformly on $\bA_i$. Let $\widehat{\tilde\pi(\bA_i)}$ denote the Monte Carlo estimate, where $\tilde\pi(\bA_i)$ is $\pi(\bA_i)$ without the normalizing constant.
It has been shown in \ctn{Bhatta21} that for perfect sampling it is enough to consider $\widehat{\tilde\pi(\bA_i)}$, instead of the true quantities
$\tilde\pi(\bA_i)$.

For $\btheta\in\bA_i$, for any Borel set $\mathbb B$ in the Borel $\sigma$-field of $\mathbb R^d$, 
let $P_i(\btheta,\mathbb B\cap\bA_i)$ denote the corresponding Metropolis-Hastings transition probability for $\pi_i$. 
Also let $Q_i(\mathbb B\cap\bA_i)$ denote the uniform distribution on $\bA_i$ with density 
\begin{equation}
	q_i(\btheta)=\frac{1}{\mathcal L(\bA_i)}I_{\bA_i}(\btheta),
	\label{eq:uniform_density}
\end{equation}
$\mathcal L(\bA_i)$ denoting the Lebesgue measure of $\bA_i$. The expression for the Lebesgue measure is analytically available and provided in \ctn{Bhatta21}.

Now let $\hat s_i$ and $\hat S_i$ denote the minimum and maximum of
$\tilde\pi(\cdot)$ over the Monte Carlo samples drawn uniformly from $\bA_i$ in the course of estimating $\tilde\pi(\bA_i)$ by $\widehat{\tilde\pi(\bA_i)}$.
Let $\hat p_i=\frac{\hat s_i}{\hat S_i}-\eta_i$, where $\eta_i$ is a sufficiently small positive quantity.
We shall refer to $\hat p_i$ as the minorization probability for $\pi_i$.
Then for all $\btheta\in\bA_i$, 
$P_i(\btheta,\mathbb B\cap\bA_i)\geq \hat p_i~ Q_i(\mathbb B\cap\bA_i)$ is the minorization inequality 
and
\begin{equation}
	R_i(\btheta,\mathbb B\cap\bA_i)=\frac{P_i(\btheta,\mathbb B\cap\bA_i)-\hat p_i~ Q_i(\mathbb B\cap\bA_i)}{1-\hat p_i}
	\label{eq:split2}
\end{equation}
is the residual distribution.
Perfect sampling from $\pi_i$ proceeds via the following steps
\begin{itemize}
\item[(a)] Draw $T_i\sim Geometric(\hat p_i)$ with respect to the mass function
\begin{equation*}
P(T_i=t)=\hat p_i (1-\hat p_i)^{t-1};~t=1,2,\ldots.
\end{equation*}
\item[(b)] Draw $\btheta^{(-T_i)}\sim Q_i(\cdot)$.
\item[(c)] Using $\btheta^{(-T_i)}$ as the initial value, carry the chain $\btheta^{(t+1)}\sim R_i(\btheta^{(t)},\cdot)$ 
forward for $t=-T_i,-T_i+1,\ldots,-1$. 
\item[(d)] Report $\btheta^{(0)}$ as a perfect realization from $\pi_i$. 
\end{itemize}
The method of simulating from $R_i(\btheta^{(t)},\cdot)$ is detailed in \ctn{Bhatta21}.
The complete algorithm for $iid$ sampling is provided as Algorithm 1 in Section 4 of \ctn{Bhatta21}.

\subsection{The role of diffeomorphism}
\label{subsec:diffeo}

As is clear from the perfect sampling step (a) following (\ref{eq:split2}), small values of $\hat p_i$ would lead to large values of $T_i$, resulting in
inefficient perfect sampling algorithm. To ensure substantially large $\hat p_i$, \ctn{Bhatta21} exploited the inverse of a diffeomorphism proposed
in \ctn{Johnson12a} to flatten
the posterior distribution in a way that its infimum and the supremum are reasonably close (so that $\hat p_i$ are adequately large) on all the $\bA_i$.

In a nutshell, if $\pi$, the multivariate target density of some random vector $\btheta$ is of interest, then  
\begin{align}
\pi_{\bgamma}(\bgamma)=\pi\left(h(\bgamma)\right)\left|\mbox{det}~\nabla h(\bgamma)\right|
\label{eq:transformed_target}
\end{align}
is the density of $\bgamma=h^{-1}(\btheta)$,
where $h$ is a diffeomorphism.
In the above, $\nabla h(\bgamma)$ denotes the gradient of $h$ at $\bgamma$ and $\mbox{det}~\nabla h(\bgamma)$ stands for the determinant of the gradient of $h$ at $\bgamma$.

\ctn{Johnson12a} obtain conditions on $h$ which make $\pi_{\bgamma}$ super-exponentially light.
Specifically, they define the following isotropic function $h:\mathbb R^d\mapsto\mathbb R^d$:
\begin{equation}
	h(\bgamma)=\left\{\begin{array}{cc}f(\|\bgamma\|)\frac{\bgamma}{\|\bgamma\|}, & \bgamma\neq \bzero\\
		0, & \bgamma=\bzero,
\end{array}\right.
\label{eq:isotropy}
\end{equation}
for some function $f: (0,\infty)\mapsto (0,\infty)$, $\|\cdot\|$ being the Euclidean norm.
\ctn{Johnson12a} confine attention to isotropic diffeomorphisms, that is, functions of the form $h$ where 
both $h$ and $h^{-1}$ are continuously differentiable, with the further property that 
$\mbox{det}~\nabla h$ and  $\mbox{det}~\nabla h^{-1}$ are also continuously 
differentiable. In particular, if $\pi$ is only sub-exponentially light, then the following form of $f:[0,\infty)\mapsto [0,\infty)$ given by
\begin{equation}
f(x)=\left\{\begin{array}{cc}e^{bx}-\frac{e}{3}, & x>\frac{1}{b}\\
x^3\frac{b^3e}{6}+x\frac{be}{2}, & x\leq \frac{1}{b},
\end{array}\right.
\label{eq:diffeo2}
\end{equation}
where $b>0$, ensures that the transformed density $\pi_{\bgamma}$ of the form (\ref{eq:transformed_target}),
is super-exponentially light.

For our purpose, the target distribution $\pi$ needs to be converted to some thick-tailed distribution $\pi_{\bgamma}$ to ensure that the supremum
and infimum of $\pi_{\bgamma}$ are close, so that $\hat p_i$ are significantly large. 
Hence, we apply the transformation $\bgamma=h(\btheta)$, the inverse of the transformation considered in \ctn{Johnson12a}.  
Consequently, the density of $\bgamma$ becomes
\begin{align}
	\pi_{\bgamma}(\bgamma)=\pi\left(h^{-1}(\bgamma)\right)\left|\mbox{det}~\nabla h(\bgamma)\right|^{-1},
\label{eq:transformed_target2}
\end{align}
where $h$ is the same as (\ref{eq:isotropy}) and $f$ is given by (\ref{eq:diffeo2}).
We also give the same transformation to the uniform proposal density (\ref{eq:uniform_density}), so that the new proposal density now becomes
\begin{equation}
	q_i(\bgamma)=\frac{1}{\mathcal L(\bA_i)}I_{\bA_i}(h^{-1}(\bgamma))\left|\mbox{det}~\nabla h(\bgamma)\right|^{-1}.
	\label{eq:proposal2}
\end{equation}
With (\ref{eq:proposal2}) as the proposal density for (\ref{eq:transformed_target2}), the proposal will not cancel in the acceptance ratio
of the Metropolis-Hastings acceptance probability. For any set $\bA$, let $h(\bA)=\left\{h(\btheta):\btheta\in\bA\right\}$.
Also, now let $s_i=\underset{\bgamma\in h(\bA_i)}{\inf}~\frac{\tilde\pi_{\bgamma}(\bgamma)}{q_i(\bgamma)}$ and 
$S_i=\underset{\bgamma\in h(\bA_i)}{\sup}~\frac{\tilde\pi_{\bgamma}(\bgamma)}{q_i(\bgamma)}$, where $\tilde\pi_{\bgamma}(\bgamma)$ is the same as (\ref{eq:transformed_target2})
but without the normalizing constant.
Then, with (\ref{eq:proposal2}) as the proposal density, we have 
\begin{align}
	P_i(\bgamma,h(\mathbb B\cap\bA_i))&\geq\int_{h(\mathbb B\cap\bA_i)}
	\min\left\{1,\frac{\tilde\pi_{\bgamma}(\bgamma')/q_i(\bgamma')}{\tilde\pi_{\bgamma}(\bgamma)/q_i(\bgamma)}\right\}q_i(\bgamma')d\bgamma'\notag\\
	&\geq p_i~Q_i(h(\mathbb B \cap\bA_i)),\notag
\end{align}
where $p_i=s_i/S_i$ and $Q_i$ is the probability measure corresponding to (\ref{eq:proposal2}).
With $\hat p_i=\hat s_i/\hat S_i-\eta_i$, where $\hat s_i$ and $\hat S_i$ are Monte Carlo estimates of $s_i$ and $S_i$ and $\eta_i>0$ is adequately small, 
the rest of the details remain the same as before with necessary modifications pertaining to the new proposal density (\ref{eq:proposal2}) and the new 
Metropolis-Hastings acceptance ratio
with respect to (\ref{eq:proposal2}) incorporated in the subsequent steps. Once $\bgamma$ is generated from (\ref{eq:transformed_target2}) we
transform it back to $\btheta$ using $\btheta=h^{-1}(\bgamma)$.

\section{IID sampling from multimodal distributions}
\label{sec:multimodal}

For any general multimodal target distribution $\pi$, using TMCMC or otherwise it is possible to identify the modes of the target distribution. 
Discussion of suitable methods for this purpose is provided in Section \ref{sec:modes}.

Let $\tilde\bmu_j$; $j=1,\ldots,m$ denote the $m$ modes of $\pi$. Consider the modal regions of the form 
$\bB_j=\left\{\btheta:\|\btheta-\tilde\bmu_j\|<\epsilon_j\right\}$,
for some sufficiently small $\epsilon_j>0$; $j=1,\ldots,m$. Let $\tilde p_j$ be the proportion of TMCMC realizations falling in $\bB_j$.  
Note that $\sum_{j=1}^m\tilde p_j=1$.
Let $\tilde\bSigma_j$ denote the empirical covariance of the TMCMC realizations falling in $\bB_j$.

For each mode $\tilde\mu_j$ and the associated covariance $\tilde\bSigma_j$, consider the infinite sequence of ellipsoids 
$\tilde\bA_{ij}=\left\{\btheta:c_{i-1,j}\leq (\btheta-\tilde\bmu_j)^T\tilde\bSigma^{-1}_j(\btheta-\tilde\bmu_j)\leq c_{i,j}\right\}$; $i=1,2,\ldots$,
where $0=c_{0,j}<c_{1,j}<c_{2,j}<\cdots$.

To simulate from $\pi$, we select $\left\{\tilde\bA_{ij}:i\geq 1\right\}$ with probability $\tilde p_j$ and apply our perfect sampling methodology
to generate exactly from $\pi$. In other words, for each $j=1,\ldots,m$, with probability $\tilde p_j$, we decompose $\pi$ as
\begin{equation*}
	\pi(\btheta)=\sum_{i=1}^{\infty}\pi(\bA_{ij})\pi_{ij}(\btheta),
\end{equation*}
where 
\begin{equation*}
	\pi_{ij}(\btheta)=\frac{\pi(\btheta)}{\pi(\bA_{ij})}I_{\bA_{ij}}(\btheta), 
\end{equation*}
and $\pi(\bA_{ij})=\int_{\bA_{ij}}\pi(d\btheta)\geq 0$. Clearly, $\sum_{i=1}^{\infty}\pi(\bA_{ij})=1$, for $j=1,\ldots,m$.
Let $\widehat{\tilde\pi(\bA_{ij})}$ stand for the Monte Carlo estimate of $\tilde\pi(\bA_{ij})$, where $\tilde\pi(\bA_{ij})$ is $\pi(\bA_{ij})$ 
without the normalization constant.

Now, for any Borel set $\bA$ we recommend Monte Carlo estimation of $\tilde\pi(\bA)$ using the transformation 
$\btheta = h^{-1}(\bgamma)$ by first noting the following:
\begin{align}
\frac{1}{\mathcal L(\bA)}\int_{\bA}\tilde\pi(\btheta)d\btheta
	&=\frac{1}{\mathcal L(\bA)}\int\tilde\pi(\btheta)I_{\bA}(\btheta)d\btheta\notag\\
	&=\frac{1}{\mathcal L(\bA)}\int\tilde\pi\left(h^{-1}(\bgamma)\right)I_{\bA}\left(h^{-1}(\bgamma)\right)\left|\mbox{det}~\nabla h(\bgamma)\right|^{-1}d\bgamma\notag\\
	&=\int\tilde\pi\left(h^{-1}(\bgamma)\right)q\left(\bgamma\right)d\bgamma\notag\\
	&=\int\tilde\pi\left(h^{-1}(\bgamma)\right)\left|\mbox{det}~\nabla h(\bgamma)\right|^{-1}
	\times\left|\mbox{det}~\nabla h(\bgamma)\right|q\left(\bgamma\right)d\bgamma\notag\\
	&=\int\tilde\pi_{\bgamma}(\bgamma)\left|\mbox{det}~\nabla h(\bgamma)\right|q\left(\bgamma\right)d\bgamma\notag\\
	&=E_q\left[\tilde\pi_{\bgamma}(\bgamma)\left|\mbox{det}~\nabla h(\bgamma)\right|\right].\label{eq:MC_diffeo}
\end{align}
In the above, $\tilde\pi_{\bgamma}(\bgamma)$ is $\pi_{\bgamma}(\bgamma)$ given by (\ref{eq:transformed_target2}) without the normalizing constant and
$q\left(\bgamma\right)$ is of the same form as (\ref{eq:proposal2}). In (\ref{eq:MC_diffeo}), $E_q$ stands for the expectation with respect to $q$.

If $\bgamma^{(\ell)}$; $\ell=1,\ldots,N$, for sufficiently large $N$ are $iid$ realizations from $q$, then the Monte Carlo estimate 
of $\tilde\pi(\bA)$ follows from (\ref{eq:MC_diffeo}) and is given by 
\begin{equation}
\widehat{\tilde\pi(\bA)}=\mathcal L(\bA)\times 
N^{-1}\sum_{\ell=1}^N\tilde\pi_{\bgamma}\left(\bgamma^{(\ell)}\right)\left|\mbox{det}~\nabla h\left(\bgamma^{(\ell)}\right)\right|.
\label{eq:MC_diffeo2}
\end{equation}
The reason for recommending the diffeomorphism based Monte Carlo estimate (\ref{eq:MC_diffeo2}) is that $\tilde\pi_{\bgamma}$ is rendered flatter
than the original target $\tilde\pi$ while $\mbox{det}~\nabla h\left(\bgamma\right)$ is a function of the one-dimensional scalar quantity $\|\bgamma\|$ (see \ctn{Johnson12a}), 
and hence is expected to lead to a more stable and reliable estimate.

\subsection{The complete algorithm for $iid$ sample generation from multimodal $\pi$}
\label{sec:complete_algo}

For given $j\in\{1,\ldots,m\}$, with respect to $\left\{\bA_{ij}:i\geq 1\right\}$, let $T_{ij}$, $\hat p_{ij}$, $\eta_{ij}$, $Q_{ij}$ and $R_{ij}$ 
stand for the analogues of $T_i$, $\hat p_i$, $\eta_i$, $Q_i$ and $R_i$, respectively.
With these notation, we present the complete algorithm for generating $iid$ realizations from the multimodal target distribution $\pi$ as Algorithm \ref{algo:perfect},
assuming the diffeomorphism based setups presented in Section \ref{subsec:diffeo} and (\ref{eq:MC_diffeo}).
\begin{algo}\label{algo:perfect}\topline
IID sampling from multimodal target distributions \botline \normalfont \ttfamily
\begin{itemize}
	\item[(1)] Using TMCMC or otherwise, obtain $\tilde\bmu_j$ and $\tilde\bSigma_j$ required for the sets $\bA_{ij}$; $i\geq 1$ and $j=1,\ldots,m$.
	\item[(2)] Fix $M$ to be sufficiently large. 
	\item[(3)] Choose the radii $\sqrt{c_{i,j}}$; $i=1,\ldots,M$; $j=1,\ldots,m$ appropriately. 
		The general strategy discussed in \ctn{Bhatta21} is adequate here and $c_{ij}$ can often be treated as the same for all $j=1,\ldots,m$.
	\item[(4)] Compute the Monte Carlo estimates $\widehat{\tilde\pi(\bA_{ij})}$; $i=1,\ldots,M$; $j=1,\ldots,m$, in parallel processors. 
	\item[(5)] Instruct each processor to send its respective estimate to all the other processors.
	\item[(6)] Let $K$ be the required $iid$ sample size from the target distribution $\pi$. Split the job of obtaining $K$ $iid$ realizations into 
		parallel processors, each processor scheduled to simulate a single realization at a time. 
		In each processor, do the following:
		\begin{enumerate}	
			\item[(i)] Select $\left\{\bA_{ij}:i\geq 1\right\}$ with probability $\tilde p_j$.
			\item[(ii)] Select $\pi_{ij}$ with probability proportional to $\widehat{\tilde\pi(\bA_{ij})}$. 
			\item[(iii)]If $i=M$ for any processor, for given $j\in\{1,\ldots,m\}$, then return to Step (2), increase $M$ to $2M$, and repeat the subsequent steps
				(in Step (4) only $\widehat{\tilde\pi(\bA_{ij})}$; $i=M+1,\ldots,2M$, need to be computed).		
				Else
				\begin{enumerate}
					\item[(a)] Draw $T_{ij}\sim Geometric(\hat p_{ij})$ with respect to 
\begin{equation*}
	P(T_{ij}=t)=\hat p_{ij} (1-\hat p_{ij})^{t-1};~t=1,2,\ldots.
\end{equation*}
					\item[(b)] Draw $\btheta^{(-T_{ij})}\sim Q_{ij}(\cdot)$.
					\item[(c)] Using $\btheta^{(-T_{ij})}$ as the initial value, carry the chain $\btheta^{(t+1)}\sim R_{ij}(\btheta^{(t)},\cdot)$ 
						forward for $t=-T_{ij},-T_{ij}+1,\ldots,-1$. 
                \item[(d)] From the current processor, send $\btheta^{(0)}$ to processor $0$ as a perfect realization from $\pi$.
				\end{enumerate}
		\end{enumerate}
	\item[(7)] Processor $0$ stores the $K$ $iid$ realizations $\left\{\btheta^{(0)}_1.\ldots,\btheta^{(0)}_K\right\}$ thus generated from the target distribution $\pi$.
\end{itemize}
\rmfamily
\botline
\end{algo}

\section{Obtaining the modes of desired distributions}
\label{sec:modes}

In this section we touch upon two different concepts leading to theories and methods of identification of the modes of the target distribution.
One such concept, elucidated in Section \ref{subsec:centrality}, relies upon the idea of centrality of the quantity of interest. The idea has been
introduced in \ctn{Sabya11} and also utilized in \ctn{Das19}.
The other, introduced by \ctn{Roy20} and discussed in Section \ref{subsec:gp}, considers embedding the objective function in a Gaussian process 
based Bayesian framework, along with the available first and second derivatives, and obtains posterior solutions that emulate the function optima. 

\subsection{Identification of modes using the concept of centrality}
\label{subsec:centrality}

Motivated by \ctn{Sabya11} who propose a methodology for obtaining the modes and any desired
highest posterior density credible regions associated with the posterior distribution of clusterings,
in Section S-7 of their supplement, \ctn{Das19} propose the same for the posterior distributions of densities. 
Here, for our purpose, we adopt the ideas of the aforementioned works to obtain the modes of target distributions on Euclidean spaces. We begin with the definition
of a central value in this regard. 

\begin{definition}
\label{def:def1}
A vector $\btheta_0$ is ``central" with respect to the probability measure $P$ which, for any $\epsilon>0$ satisfies the following equation:
\begin{align}
	P\left(\left\{\btheta:\|\btheta_0-\btheta\|<\epsilon\right\}\right)=\sup_{\bzeta}P\left(\left\{\btheta:\|\bzeta-\btheta\|<\epsilon\right\}\right).
\label{eq:central}
\end{align}
\end{definition}

Observe that $\btheta_0$ is the global mode of the distribution as $\epsilon\rightarrow 0$.
If the distribution is unimodal, then the central vector remains the same for all $\epsilon>0$.
However, for multimodal distributions, the central vector varies with $\epsilon$, signifying 
existence of local modes, which we define as follows.

\begin{definition}
\label{def:def2}
We define $\tilde\btheta$ to be a local mode if
\begin{align}
\lim_{\epsilon\downarrow 0}\frac{\sup_{\bzeta\in\mathcal N(\tilde\btheta,\eta)}
P\left(\left\{\btheta\in\mathcal N(\tilde\btheta,\eta):\|\bzeta-\btheta\|<\epsilon\right\}\right)}
{P\left(\left\{\btheta\in\mathcal N(\tilde\btheta,\eta):\|\tilde\btheta-\btheta\|<\epsilon\right\}\right)}
&=1,\notag
\end{align}
where $\mathcal N(\tilde\btheta,\eta)=\left\{\btheta:\|\tilde\btheta-\btheta\|<\eta\right\}$ for some $\eta>0$.
\end{definition}


Note that the central value $\btheta_0$ given by (\ref{eq:central}) is not analytically available and empirical methods
based available TMCMC realizations, are necessary.
In this regard, we consider the following definition of approximately central value.

\begin{definition}
\label{def:def3}
Assume that TMCMC realizations $\{\btheta^{(j)};~j=1,\ldots,N\}$ are available.
We define that $\btheta^{(j)}$ as ``approximately central" which, for a given small $\epsilon>0$, satisfies 
the following equation:
\begin{equation}
\btheta^{(j)}=\arg\max_{1\leq i\leq N}\frac{1}{N}\#\left\{\btheta^{(\ell)};1\leq \ell\leq N:\|\btheta^{(i)}-\btheta^{(\ell)}\|<\epsilon\right\},
\label{eq:empirical_central_density}
\end{equation}
where, for any discrete set $\bB$, $\#\bB$ stands for the number of elements in $\bB$. 
\end{definition}
The approximate central value $\btheta^{(j)}$ is easily computable and the ergodic theorem ensures convergence
of $\btheta^{(j)}$ almost surely to the true central value $\btheta_0$, as $N\rightarrow\infty$.
Varying $\epsilon$ in (\ref{eq:empirical_central_density}) enables identification of the local modes. 
In practice, we shall re-scale the Euclidean distance between the TMCMC realizations by the maximum distance with respect to the simulated realizations,
so that $\epsilon$ takes values in $(0,1)$.

\subsection{Identification of modes by function optimization with posterior Gaussian derivative process}
\label{subsec:gp}

On a rigorous footing, \ctn{Roy20} develop a novel and general Bayesian algorithm for optimization of functions whose first and second partial derivatives are known. 
The key concept underlying their contribution is the Gaussian process representation of the function which induces a first derivative process that is also Gaussian. 
Given suitable choices of input points in the function domain and their function values that constitute the data, 
the stationary points of the objective function are emulated by Bayesian posterior solutions of the derivative process set equal to zero. 
The method is fine-tuned by setting restrictions on the prior in terms of the first and second derivatives of the objective function. 
\ctn{Roy20} demonstrate successful applications of their method in various examples, including problems involving multiple optima.

Hence, treating $\tilde\pi(\btheta)$ as the objective function to be maximized with respect to $\btheta$, the Bayesian algorithm of \ctn{Roy20}
may be employed to obtain multiple modes.


\section{Simulation experiment to demonstrate $iid$ sampling from multimodal distributions}
\label{sec:simstudy_multmodal}

\subsection{The setup}
\label{subsec:setup}
To illustrate our $iid$ sampling methodology, we now apply the same to generate $iid$ realizations from 
a mixture of two $50$-dimensional normal distributions. To specify the normal mixture, with $d=50$, let us first set
$\bnu=(\nu_1,\ldots,\nu_d)^T$, with $\nu_i=i$, for $i=1,\ldots,d$, and consider a $d\times d$ 
scale matrix $\bS$, whose $(i,j)$-th element is specified by $S_{ij}=10\times\exp\left\{-(i-j)^2/2\right\}$. 
We consider two $d$-dimensional normal components, the means being $\bnu_1=\bnu$ and $\bnu_2=2\bnu$, and the covariance matrix $\bS$ the same 
as the above for both the normal components. The mixing proportions
of the normals associated with $\bnu_1$ and $\bnu_2$ are $2/3$ and $1/3$, respectively. 

Note that the normal mixture has been used by \ctn{Bhatta21} to demonstrate $iid$ sampling, but knowledge of means and covariances of the mixture 
components and the mixing proportions are assumed, which are usually unavailable in practice. Here we demonstrate application of  
Algorithm \ref{algo:perfect} to generate $10,000$ $iid$
realizations from this mixture, without assumptions of availability of such information.  


\subsection{Implementation and results}
\label{subsec:imp_results}
First, we employed the methodology of \ctn{Roy20} to find the modes of the mixture distribution, and the results $\tilde\bmu_1$ and $\tilde\bmu_2$ 
turned out to be significantly close to $\bnu_1$ and $\bnu_2$, respectively. Now, in the modal regions $\bB_1$ and $\bB_2$ 
described in Section \ref{sec:multimodal}, we set $\epsilon_1=\epsilon_2=0.92$. This choice corresponds to the maximum values of $\epsilon_1$ and
$\epsilon_2$ such that all the probabilities $\hat p_{ij}$, for $i=1,\ldots,M=10^5$ and $j=1,2$, are non-zero. 
The corresponding mixing probability estimates turned out to be $\tilde p_1=0.6672$ and $\tilde p_2=0.3328$, which are close to the true mixing probabilities.

For implementing Algorithm \ref{algo:perfect}, we set the diffeomorphism parameter $b=0.01$, $\sqrt{c_{1,j}}=0.05$ and for $i=2,\ldots,M=10^5$; $j=1,2$, set 
$\sqrt{c_{i,j}}=\sqrt{c_{1,j}}+9.5\times 10^{-5}\times (i-1)$. For each $i$ and $j$, 
we set the Monte Carlo size to be $5000$. 
These choices are instrumental for leading
to significantly large choices of $\hat p_{ij}$. In $\hat p_{ij}$, we set $\eta_{ij}=10^{-10}$.

All our codes are written in C using the Message Passing Interface (MPI) protocol for parallel processing. We implemented our codes on a $80$-core VMWare 
provided by Indian Statistical Institute. The machine has $2$ TB memory and each core has about $2.8$ GHz CPU speed.

\begin{figure}
	\centering
	\subfigure [True and $iid$-based density for $\theta_1$.]{ \label{fig:n1}
	\includegraphics[width=7.5cm,height=7.5cm]{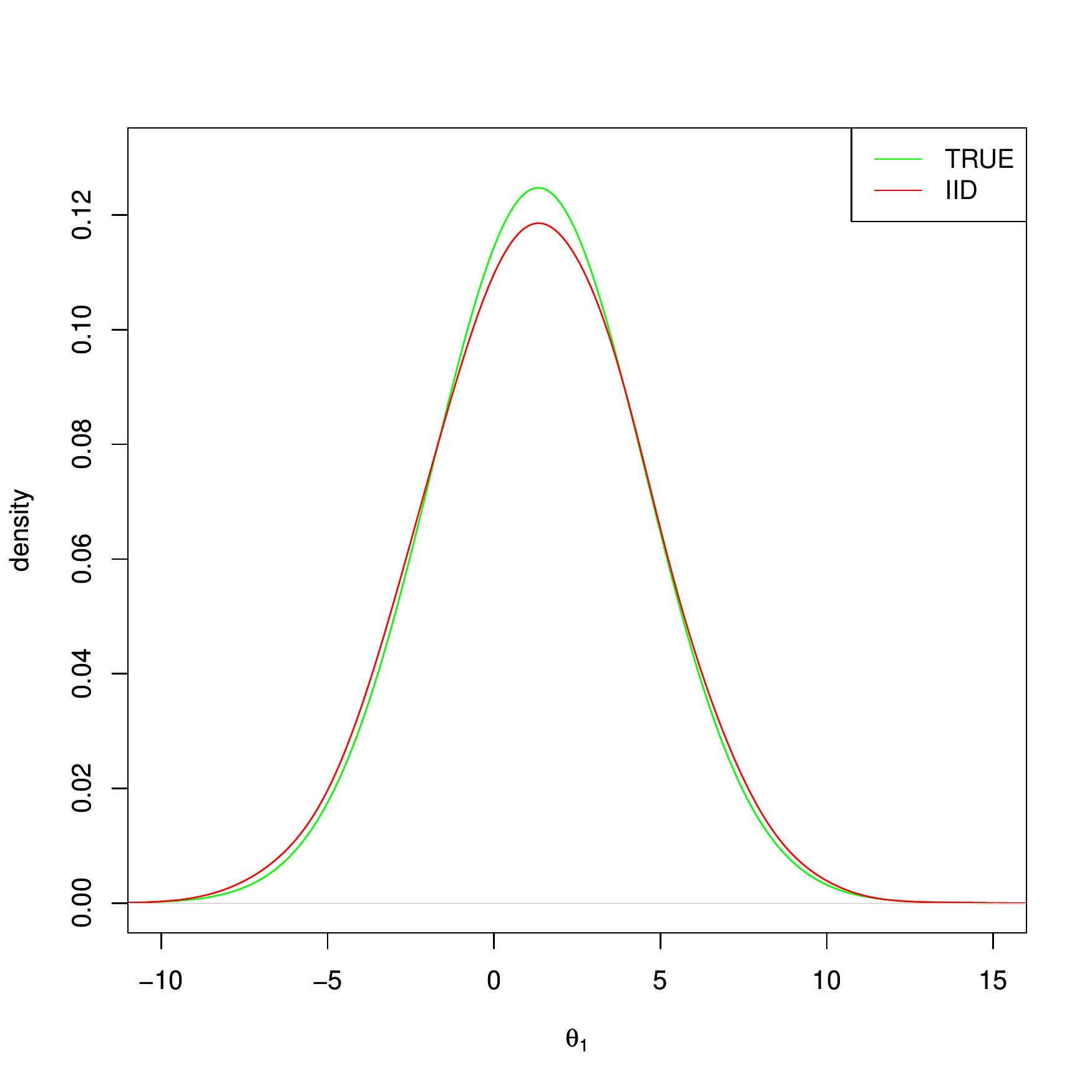}}
	\hspace{2mm}
	\subfigure [True and $iid$-based density for $\theta_{10}$.]{ \label{fig:n10}
	\includegraphics[width=7.5cm,height=7.5cm]{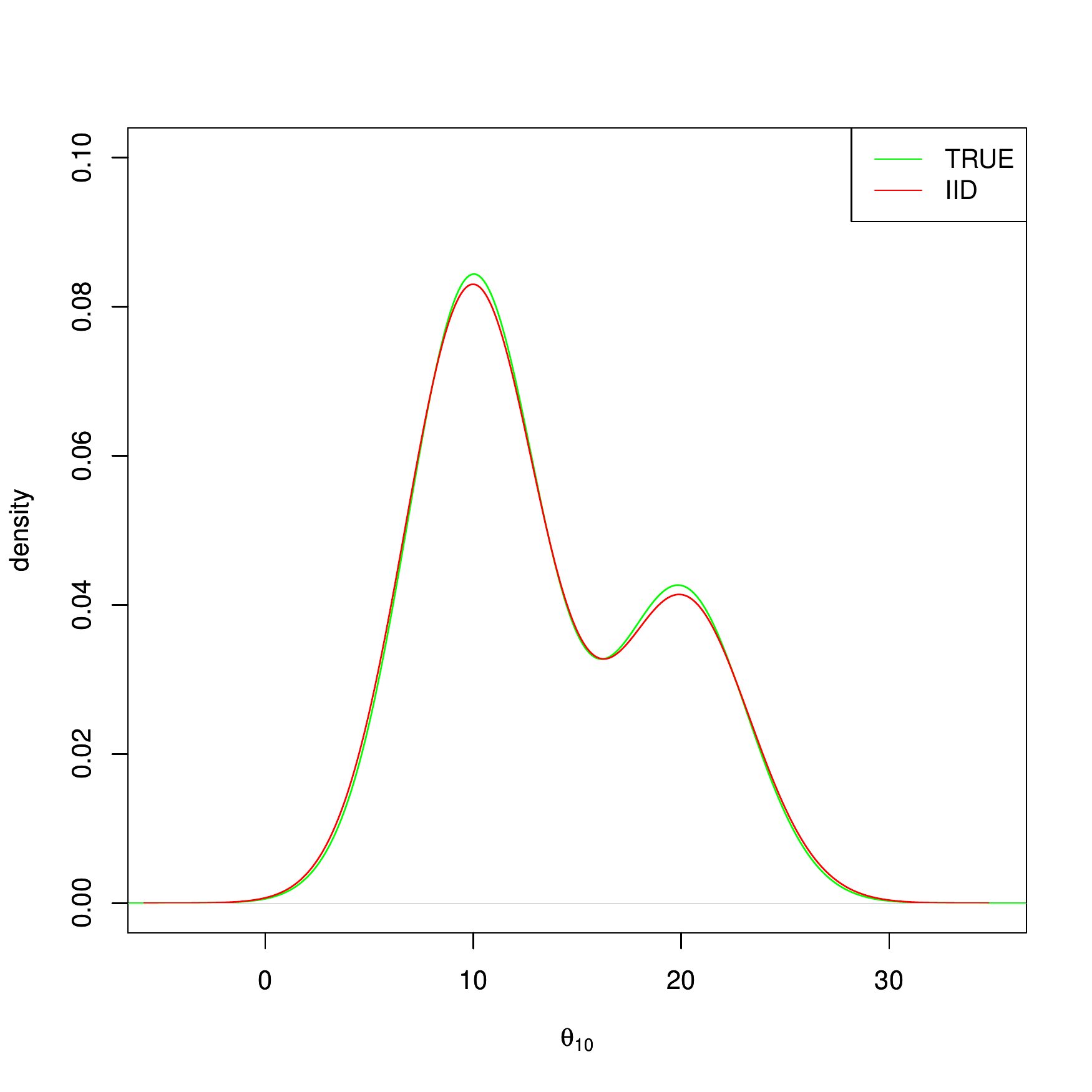}}\\
	\vspace{2mm}
	\subfigure [True and $iid$-based density for $\theta_{25}$.]{ \label{fig:n25}
	\includegraphics[width=7.5cm,height=7.5cm]{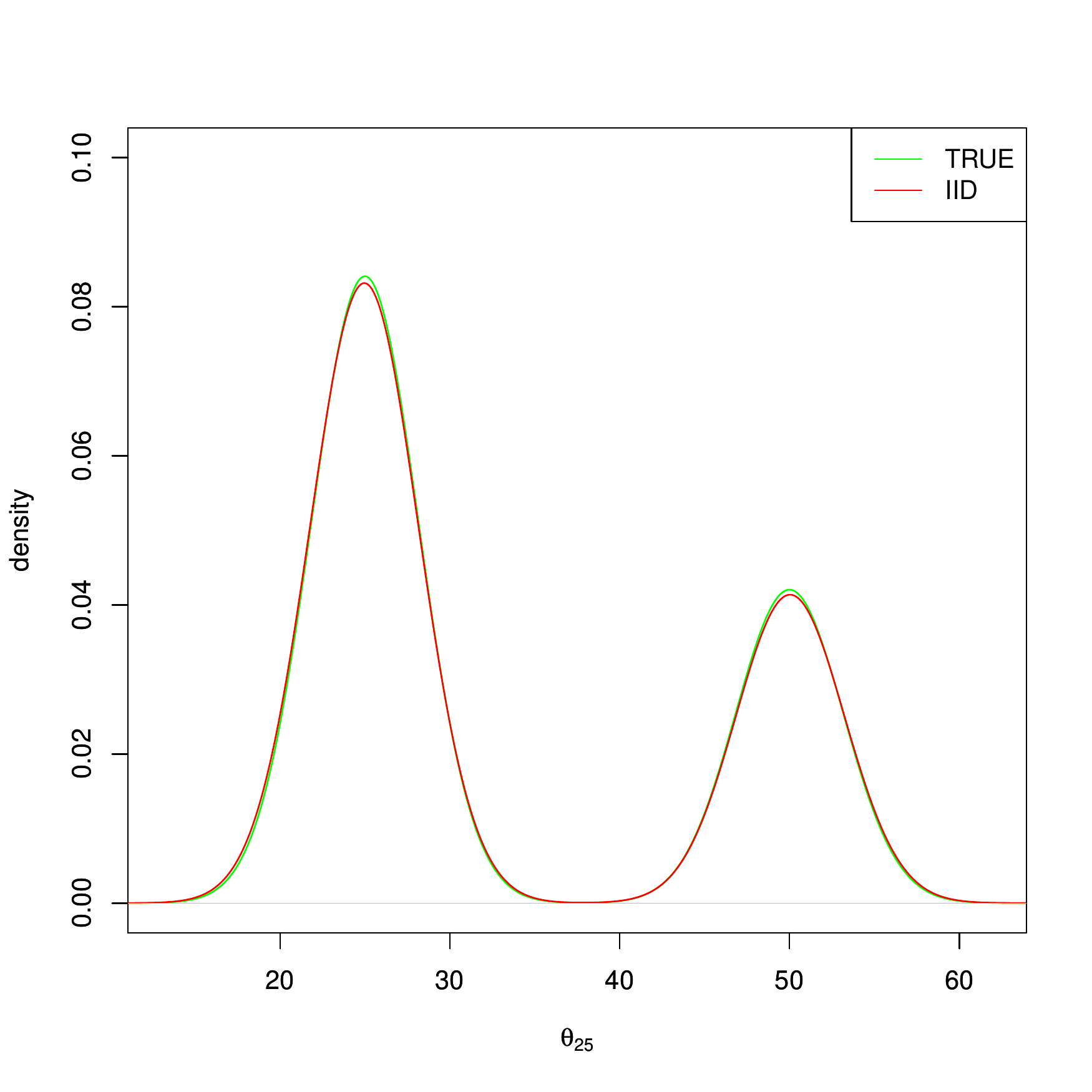}}
	\vspace{2mm}
	\subfigure [True and $iid$-based density for $\theta_{50}$.]{ \label{fig:n50}
	\includegraphics[width=7.5cm,height=7.5cm]{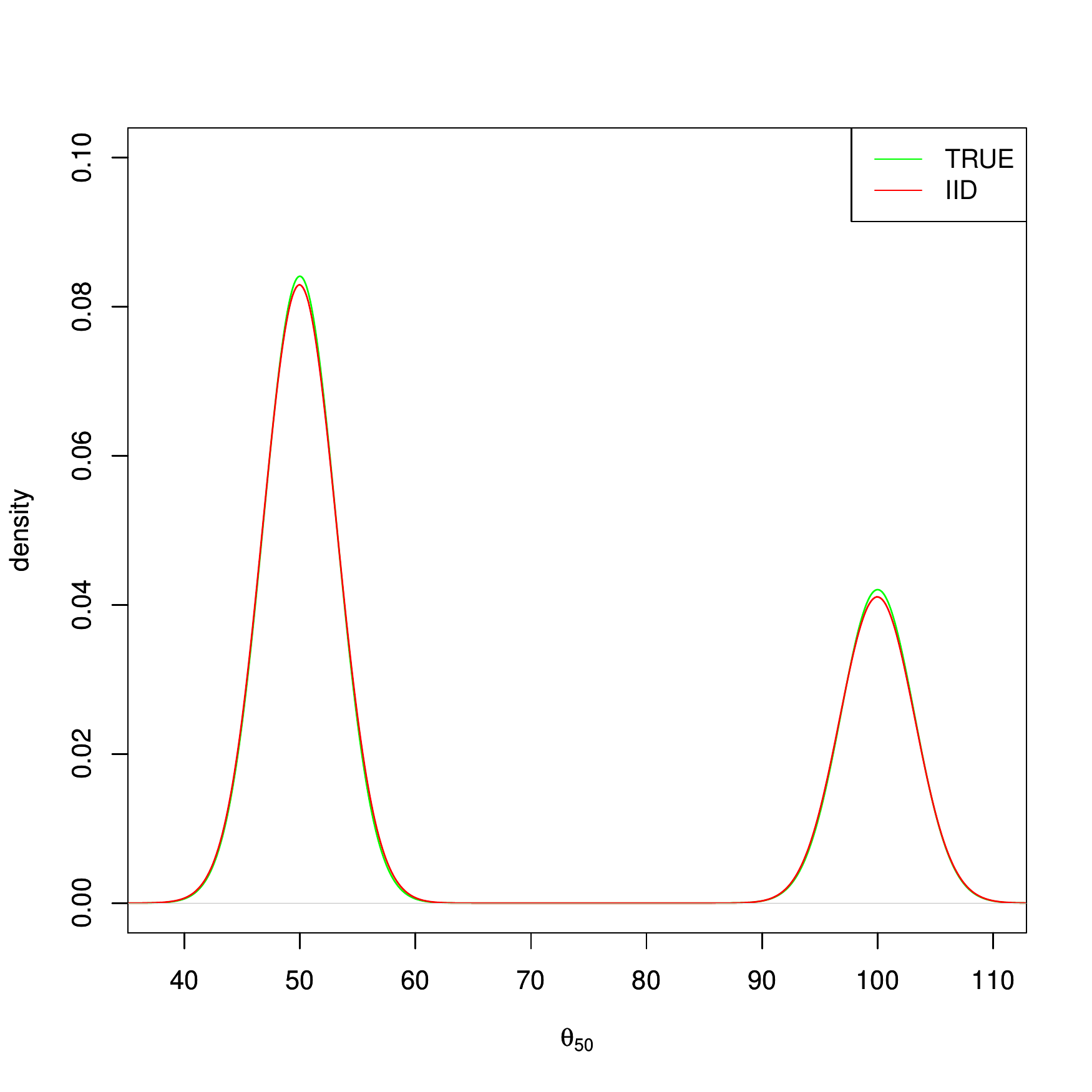}}
	\caption{Simulation from $50$-dimensional mixture normal distribution. 
	The red and green colours denote the $iid$ sample based density and the true density, respectively.}
	\label{fig:normix_simstudy50}
\end{figure}

In our implementation, it takes about an hour to generate $10,000$ $iid$ realizations from this $50$-dimensional normal mixture.
Figure \ref{fig:normix_simstudy50} vindicates quite accurate performance of our $iid$ simulation method for multimodal target distributions. 


\section{IID sampling from variable-dimensional distributions}
\label{sec:vardim}

In the realm of variable dimensions, the distribution of interest is $\pi(k,\btheta_k)$, where $k$ takes values in a set of countable indices $\mathcal I$, say, 
while $\btheta_k\in\mathbb R^{d_k}$ denotes the $d_k$-dimensional parameter. Letting $\mathcal C_k= \{k\}\times\mathbb R^{d_k}$, it is clear that
$(k,\btheta_k)$ varies over $\cup_{\ell\in\mathcal I}\mathcal C_{\ell}$.

In the Bayesian setup, the posterior distribution $\pi(k,\btheta_k|\by)$ is of interest, where $\by$ denotes observed data.
Note that $\pi(k,\btheta_k|\by)=\pi(k|\by)\pi(\btheta_k|k,\by)$.
Hence, for simulation from the variable-dimensional posterior $\pi(k,\btheta_k|\by)$, ideally one may simulate from $\pi(k|\by)$ and given the simulated value $k'$, say,
may simulate $\btheta_{k'}$ from $\pi(\btheta_{k'}|k',\by)$. However, although at least MCMC methods can be employed for (approximately) sampling from $\pi(\btheta_k|k,\by)$,
generating draws from $\pi(k|\by)$ requires first integrating out $\btheta_k$ from the joint posterior $\pi(k,\btheta_k|\by)$, for all $k\in\mathcal I$, which 
is usually considered infeasible. The latter technical issue is responsible for the birth of a plethora of variable-dimensional MCMC strategies, ranging from
the extremely inefficient and aesthetically unappealing RJMCMC to the efficient and elegant TTMCMC.

With our $iid$ sampling theory and method, we can generate exact $iid$ samples from $\pi(\btheta_k|k,\by)$, for any given $k$. Hence, if we can solve the problem
of generating $iid$ draws from $\pi(k|\by)$, then the variable-dimensional setup would reduce to simulation from fixed dimensional distributions corresponding to
$k$ with non-zero posterior probabilities.

Note that $\pi(k|\by)=C\pi(k)f(\by|k)$, where $\pi(k)$ is the prior for $k$, 
\begin{equation}
f(\by|k)=\int f(\by|\btheta_k,k)\pi(\btheta_k|k)d\btheta_k
	\label{eq:y_k}
\end{equation}
and $C=\left[\sum_{\ell\in\mathcal I}\pi(\ell)f(\by|\ell)\right]^{-1}$ is the normalization constant.
In principle, the integration problem posed by (\ref{eq:y_k}) can be certainly handled by the Monte Carlo method with realizations simulated from the prior $\pi(\btheta_k|k)$,
but the resultant estimate may be poor, since a large proportion of the prior based realizations may represent regions where $f(\by|\btheta_k,k)$ is negligibly small.

Let $\bmu_k$ and $\bSigma_k$ denote the mean (or mode) and covariance (or appropriate scale matrix) 
of $\pi(\btheta_k|k,\by)$, for $k\in\mathcal I$. 
%
For each $\bmu_k$ and $\bSigma_k$, consider the infinite sequence of ellipsoids 
$\bA_{ik}=\left\{\btheta_k:c_{i-1,k}\leq (\btheta_k-\bmu_k)^T\bSigma^{-1}_k(\btheta_k-\bmu_k)\leq c_{i,k}\right\}$; $i=1,2,\ldots$,
where $0=c_{0,k}<c_{1,k}<c_{2,k}<\cdots$. As before, this sequence will be used to simulate $iid$ samples from $\pi(\btheta_k|k,\by)$, for any $k\in\mathcal I$ that
is supported by $\pi(k|\by)$.
However, this sequence has a further importance: it may be used for the purpose of reliable approximation of the integral (\ref{eq:y_k}).
Indeed, writing $\tilde\pi_k(\btheta_k)=f(\by|\btheta_k,k)\pi(\btheta_k|k)$, note that
\begin{align}
	f(\by|k)&=\int \tilde\pi(\btheta_k)d\btheta_k\notag\\
	&=\sum_{i=1}^{\infty}\mathcal L(\bA_{ik})\int\tilde\pi_k(\btheta_k)\frac{I_{\bA_{ik}}(\btheta_k)}{\mathcal L(\bA_{ik})}d\btheta_k\notag\\
	&=\sum_{i=1}^{\infty}\mathcal L(\bA_{ik})E_{q_{ik}}\left[\tilde\pi_{\bgamma_k}(\bgamma_k) \left|\mbox{det}~\nabla h(\bgamma_k)\right|\right],
	\label{eq:MC_diffeo3}
\end{align}
where the expression within the summation of (\ref{eq:MC_diffeo3}) follows in the same way as (\ref{eq:MC_diffeo}), with 
$\btheta_k = h^{-1}(\bgamma_k)$, $\tilde\pi_{\bgamma_k}(\bgamma_k)=\tilde\pi_k\left(h^{-1}(\bgamma_k)\right) \left|\mbox{det}~\nabla h(\bgamma_k)\right|^{-1}$,
\begin{equation}
	q_{ik}(\bgamma)=\frac{1}{\mathcal L(\bA_{ik})}I_{\bA_{ik}}(h^{-1}(\bgamma))\left|\mbox{det}~\nabla h(\bgamma)\right|^{-1}.
	\label{eq:proposal3}
\end{equation}
and $E_{q_{ik}}\left[\tilde\pi_{\bgamma_k}(\bgamma_k) \left|\mbox{det}~\nabla h(\bgamma_k)\right|\right]$ is the expectation of 
$\tilde\pi_{\bgamma_k}(\bgamma_k) \left|\mbox{det}~\nabla h(\bgamma_k)\right|$ with respect to (\ref{eq:proposal3}).

For each $i\geq 1$, $E_{q_{ik}}\left[\tilde\pi_{\bgamma_k}(\bgamma_k) \left|\mbox{det}~\nabla h(\bgamma_k)\right|\right]$ is expected to be reliably estimated
by the Monte Carlo average of $\tilde\pi_{\bgamma_k}(\bgamma_k) \left|\mbox{det}~\nabla h(\bgamma_k)\right|$ with respect to realizations drawn from (\ref{eq:proposal3}),
due to the relatively flat structure of diffeomorphism based $\tilde\pi_{\bgamma_k}(\bgamma_k)$ and the narrow regions $\bA_{ik}$ on which uniform samples are generated
for the Monte Carlo purpose. 

For each $k$, Monte Carlo based estimation of (\ref{eq:MC_diffeo3}) is a highly parallelisable exercise: computation of $\mathcal L(\bA_{ik})$ and
estimation of $E_{q_{ik}}\left[\tilde\pi_{\bgamma_k}(\bgamma_k) \left|\mbox{det}~\nabla h(\bgamma_k)\right|\right]$ can be performed simultaneously for all $i\geq 1$ in 
parallel processors, and the results can then be combined into the sum in a straightforward manner. In practice, the infinite sum will of course be replaced with 
the sum of the first $M_k$ terms, where $M_k$ is so large that further terms have insignificant contribution when added to the first $M_k$ terms.

Recall that our strategy of estimating $f(\by|k)$ through (\ref{eq:MC_diffeo3}) assumes a single $\bmu_k$ and $\bSigma_k$ for the posterior
$\pi(\btheta_k|k,\by)$. In the case of multimodality,
we shall set $\bmu_k$ to be the global mode obtained by the methods discussed in Section \ref{sec:modes} and $\bSigma_k$ would be the empirical covariance
of the TMCMC realizations from $\pi(\btheta_k|k,\by)$ falling in $\bB_k=\left\{\btheta_k:\|\btheta_k-\bmu_k\|<\epsilon_k\right\}$, for adequately small $\epsilon_k$. 

Assuming that the posteriors $\pi(\btheta_k|k,\by)$ may be multimodal for some or all $k$ with $m_k$ ($\geq 1$) modes, $iid$ sampling from variable-dimensional distributions 
can be summarized by the algorithm below.
\begin{algo}\label{algo:perfect2}\topline
IID sampling from variable-dimensional target distributions \botline \normalfont \ttfamily
\begin{itemize}
	\item[(1)] Using TMCMC or otherwise, obtain the modes $\tilde\bmu_{jk}$ and the corresponding covariance matrices $\tilde\bSigma_{jk}$ 
		required for the sets 
		$$\bA_{ijk}=\left\{\btheta_k:c_{i-1,j,k}\leq (\btheta_k-\tilde\bmu_{jk})^T\tilde\bSigma^{-1}_{jk}(\btheta_k-\tilde\bmu_{jk})\leq c_{i,j,k}\right\};$$ 
		$i=1,2,\ldots$,
where $0=c_{0,j,k}<c_{1,j,k}<c_{2,j,k}<\cdots$, for $i\geq 1$ 
		and $j=1,\ldots,m_k$, for $k\in\mathcal I$. Let the sequence of sets $\bA_{ik}$ be the same as the sequence $\bA_{ijk}$ for some fixed $j$ depending upon $k$, 
		corresponding to the global mode and the corresponding covariance matrix.  
	\item[(2)] For each $k$, fix $M_k$ to be sufficiently large. 
		For each $i\in\{1,\ldots,M_k\}$, compute $\mathcal L(\bA_{ik})$ and estimate 
		$E_{q_{ik}}\left[\tilde\pi_{\bgamma_k}(\bgamma_k) \left|\mbox{det}~\nabla h(\bgamma_k)\right|\right]$ 
		by Monte Carlo averaging in independent parallel processors, and finally combine the results as the summation of the $M_k$ terms. This constitutes an estimate of 
		$f(\by|k)$ given by (\ref{eq:MC_diffeo3}). Let $\hat f(\by|k)$ denote the estimate.
	\item[(3)] Repeat Step (2) for all $k\in\mathcal I$. 
	\item[(4)] Generate $K$ $iid$ realizations from $\pi(k|\by)$, which is a multinomial distribution with probabilities proportional to $\pi(k)\hat f(\by|k)$, for
		$k\in\mathcal I$. Let $\left\{k^{(r)}:r=1,\ldots,K\right\}$ denote the $K$ $iid$ realizations.
	\item[(5)] For each $k^{(r)}$; $r=1,\ldots,K$, consider the sequence of sets $\bA_{ijk^{(r)}}$, for $i\geq 1$ and $j=1,\ldots,m_{k^{(r)}}$, 
		and apply Algorithm \ref{algo:perfect} to generate a perfect realization $\btheta^{(r)}_{k^{(r)}}$ from $\pi(\btheta_{k^{(r)}}|k^{(r)},\by)$.
	\item[(6)] Store the $K$ $iid$ realizations $\left\{\left(k^{(r)},\btheta^{(r)}_{k^{(r)}}\right):r=1,\ldots,K\right\}$ 
		as an $iid$ sample from the variable-dimensional target distribution $\pi(k,\btheta_k|\by)$.
\end{itemize}
\rmfamily
\botline
\end{algo}

\section{Illustration of variable-dimensional $iid$ sampling using acidity data}
\label{sec:vardim_example}

\ctn{Das19} illustrated TTMCMC on normal mixture models with unknown number of components
with applications to the well-studied enzyme, acidity and galaxy data sets. 
The TTMCMC results of \ctn{Das19} demonstrate that the acidity data is the simplest in the sense that the posterior distribution of $k$ supports 
two and three components only with probabilities $0.9941$ and $0.0059$, respectively. In this context, note that for our $iid$ sampling procedure, 
the larger the number of values of $k$ supported by the posterior, greater the effort necessary to generate $iid$ samples from the variable-dimensional posterior.
This is of course clear since $iid$ realizations from $\pi(\btheta_k|k,\by)$ for larger number of $k$-values would be necessary, which would require
obtaining the modes and the modal regions and Monte Carlo sampling and subsequent perfect sampling for many different dimensions.
Thus, for our purpose, here we consider the acidity data along with the model and prior setup of \ctn{Das19}. 
The data consists of $n=155$ observations in the interval $(2,8)$.
The model and prior are discussed in Sections \ref{subsec:normix} and \ref{subsec:normix_prior}.

\subsection{{\bf Normal mixture}}
\label{subsec:normix}

Let the data points of $\by=(y_1,\ldots,y_n)$ be independently and identically distributed  
as the normal mixture of the following form: for $i=1,\ldots,n$
\begin{equation}
f(y_i\vert\bnu_k,\btau_k,\bpi_k,k)=\sum_{j=1}^k\pi_j\sqrt{\frac{\tau_j}{2\pi}}\exp\left\{-\frac{\tau_j}{2}(y_i-\nu_j)^2\right\},
\label{eq:normix}
\end{equation}
where $\bnu_k=(\nu_1,\ldots,\nu_k)$, $\btau_k=(\tau_1,\ldots,\tau_k)$, and $\bpi_k=(\pi_1,\ldots,\pi_k)$.
Given $k>0$, for each $j$, $-\infty<\nu_j<\infty$, $\tau_j>0$, $0<\pi_j<1$ such that
$\sum_{j=1}^k\pi_j=1$. We assume that $k$ is unknown.

\subsection{{\bf Prior structure}}
\label{subsec:normix_prior}

Following \ctn{Das19}
we consider
the following prior for $\bnu$ and $\btau$:
\begin{align}
[\tau_j] &\sim\mathcal G\left(\frac{s}{2},\frac{S}{2}\right);\notag\\ 
[\nu_j\vert\tau_j] &\sim N\left(\nu_0,\frac{\psi}{\tau_j}\right).\notag 
\end{align}
In the above, by $\mathcal G\left(a,b\right)$ we mean a gamma distribution with mean $a/b$ and variance $a/b^2$ and 
$N\left(\mu,\sigma^2\right)$ denotes the normal distribution with mean $\mu$ and variance $\sigma^2$.
As in \ctn{Das19} we set the values of the hyperparameters to be $s=4.0$, $S=2\times(0.2/0.573)$, $\nu_0=5.02$ and $\psi=33.3$. 

Following \ctn{Das19} we reparameterize $\tau_j$
as $\exp(\tau^*_j)$, where $\tau^*_j\sim\mathcal \log\left(\mathcal G(s/2,S/2)\right)$ 
and denote $(\tau^*_1,\ldots,\tau^*_k)$ by $\btau^*_k$.

For $\bpi$ we consider the reparameterization: 
for $j=1,\ldots,k$,
\begin{align}
\pi_j=\frac{\exp\left(\omega_j\right)}{\sum_{\ell=1}^k\exp\left(\omega_j\right)};\quad
\omega_1,\ldots,\omega_k\stackrel{iid}{\sim}N\left(\mu_{\omega},\sigma^2_{\omega}\right),
\notag
\end{align}
with $\mu_{\omega}=0$ and $\sigma^2_{\omega}=0.5$. 
Let $\bomega_k=(\omega_1,\ldots,\omega_k)$.

We set $\btheta_k=(\bnu_k,\btau^*_k,\bomega_k)$. 
As regards the prior for $k$, we consider the uniform distribution on $\{1,2,\ldots,30\}$.

\subsection{Implementation of Algorithm \ref{algo:perfect2}}
\label{subsec:implementation_acidity}

\subsubsection{Computation of $\hat f(\by|k)$ and simulation from $\pi(k|\by)$}
\label{subsubsec:compute_f}
To obtain the estimates $\hat f(\by|k)$ in Algorithm \ref{algo:perfect2} we needed to obtain $\tilde\bmu_{jk}$ and $\tilde\Sigma_{jk}$, for $j=1\ldots,m_k$ and $k=1,\ldots,30$.
In this regard, we first implemented TMCMC for the posteriors $\pi(\btheta_k|k,\by)$, for $k=1,\ldots,30$, simultaneously on $30$ parallel processors.
For each $k$ we considered a total TMCMC run length of $5000\times 150+10000\times 150=2250000$ iterations, discarding the first $5000\times 150=750000$ iterations as burn-in
and subsequently storing one in $150$ iterations to obtain $10,000$ TMCMC realizations. The total time taken for all $30$ TMCMC exercises with our parallel implementation
is about $27$ minutes. Now, rather than obtaining the modes of the posteriors from the TMCMC realizations, we considered the means of the posteriors based on the
TMCMC samples, to simplify the onerous task of obtaining the modes of $30$ posteriors using the methods discussed in Section \ref{sec:modes}.
As regards the covariance, we simply took the empirical covariance based on the TMCMC samples. In this regard, the role of diffeomorphism in computing
$\hat f(\by|k)$ is important: flattening the posterior distribution with $b=0.3$ ensured that even the TMCMC based means and covariances led 
to reliable estimates of $\hat f(\by|k)$, along with the radii $\sqrt{c_{1,k}}=0.05$ and $\sqrt{c_{i,k}}=\sqrt{c_{1,k}}+3\times 10^{-5}\times (i-1)$, for $i=2,\ldots,M_k=10^5$,
for $k=1,\ldots,30$. We set the Monte Carlo size to $5000$, as before.
The numerical estimates $\hat f(\by|k)$, for $k=1,\ldots,30$, turned out to be such that $\pi(k|\by)$ associated with these gave 
full mass to $k=2$.
This is consistent with the TTMCMC result of \ctn{Das19} that 
the posterior $\pi(k|\by)$ supports $k=2$ with probability $0.9941$ (and $k=3$ with probability $0.0059$).
It is important to mention that even direct Monte Carlo estimation of $\hat f(\by|k)$ without diffeomorphism led to exactly the same result.


\subsubsection{IID sampling from the posteriors of $\btheta_k$ given $k$ and $\by$}
\label{subsubsec:iid_vardim}
Since the posterior of $k$ supports only $k=2$, 
it is sufficient to generate $iid$ samples only from $\pi(\btheta_k|k=2,\by)$. 
TMCMC realizations from $\pi(\btheta_k|k=2,\by)$ did not show any evidence of multimodality, and hence we apply Algorithm \ref{algo:perfect} with $m=1$
(which is identical to Algorithm 1 of \ctn{Bhatta21} when diffeomorphism is considered for Monte Carlo estimation and perfect sampling) 
to simulate perfectly from this posterior, with the posterior mean and covariance estimated
from the TMCMC sample required for construction of the sets $\bA_{i1}$; $i\geq 1$. We set the radii to be 
$\sqrt{c_{1,1}}=0.05$ and $\sqrt{c_{i,1}}=\sqrt{c_{1,1}}+0.000105\times (i-1)$, for $i=2,\ldots,M=10^5$, and the Monte Carlo sample size to $5000$. 
We fixed $b=0.01$ to be the diffeomorphism parameter. 
Again, these choices ensured significant values of the 
minorization probabilities $\hat p_{i1}$; $i\geq 1$, where we fixed $\eta_i=10^{-10}$ for all $i$. Note that for computing $\hat f(\by|k)$
we had chosen $b=0.3$ to be the diffeomorphism parameter value, for all $k\in\{1,\ldots,30\}$. 
However, that value failed to yield significant minorization probabilities for $\pi(\btheta_k|k,\by)$. Indeed, in general it should not be expected that
the diffeomorphism parameters for computing $\hat f(\by|k)$ and perfect sampling from $\pi(\btheta_k|k,\by)$ should be the same to get the best results
for both the exercises. In fact, the diffeomorphism parameter $b$ should ideally be dependent upon $k$ and the associated ellipsoids, 
and the dependence structures for the two aforementioned problems would likely be quite different.
Generation of $10,000$ $iid$ realizations from $\pi(\btheta_k|k=2,\by)$ took less than a minute in our implementation of
Algorithm \ref{algo:perfect}.

Now, although the posterior $\pi(\btheta_k|k=3,\by)$ is not relevant for this exercise, we still draw $iid$ samples from this, since at least in the TTMCMC
implementation of \ctn{Das19}, $\pi(k=3|\by)$ was positive. Another reason for giving importance to this posterior is that
TMCMC realizations from $\pi(\btheta_k|k=3,\by)$ displayed vivid evidence of bimodality, which makes it interesting and instructive from our exact $iid$
sampling perspective. 
As such, we first considered the centrality idea
discussed in Section \ref{subsec:centrality} to seek out the modes of the posterior. Adopting the empirical definition 
(Definition \ref{eq:empirical_central_density}) and allowing $\epsilon$ to range between $0.001$ and $0.999$ with spacing $0.1$,
we found $6$ modes of $\pi(\btheta_k|k=3,\by)$ corresponding to $\epsilon=0.101$. Implementation of the Gaussian process optimization method of \ctn{Roy20}
with these modes as initial values yielded results that failed to better the results of centrality. In other words, the $6$ modes obtained by 
Definition \ref{eq:empirical_central_density} prevailed even with respect to the rigorous theory developed in \ctn{Roy20}. We then obtained
the empirical covariance matrices and mixing probabilities corresponding to these modes following the ideas discussed in Section \ref{sec:multimodal}, with $\epsilon_j=0.33$
for $j=1,\ldots,m$, where $m=6$ in this situation. The choice $0.33$ here is the largest value making the minorizing probabilities significant, subject
to positive definiteness of the empirical covariance matrices, and the further choices
$\sqrt{c_{1,2}}=0.05$ and $\sqrt{c_{i,2}}=\sqrt{c_{1,2}}+5\times 10^{-5}\times (i-1)$, for $i=2,\ldots,M=10^5$, and $b=0.01$ as the diffeomorphism parameter. 
The mixing probabilities $\tilde p_j$; $j=1,\ldots,6$, turned out to be proportional to $0.0024$, $0.0014$, $0.0017$, $0.001$, $0.0011$ and $0.0017$.
As before, we set the Monte Carlo sample size to $5000$. It took $46$ minutes in our parallel implementation to generate $10,000$ $iid$ realizations 
from $\pi(\btheta_k|k=3,\by)$ using Algorithm \ref{algo:perfect}.

%

\begin{figure}
	\centering
	\subfigure []{ \label{fig:nu1}
	\includegraphics[width=6.0cm,height=6.0cm]{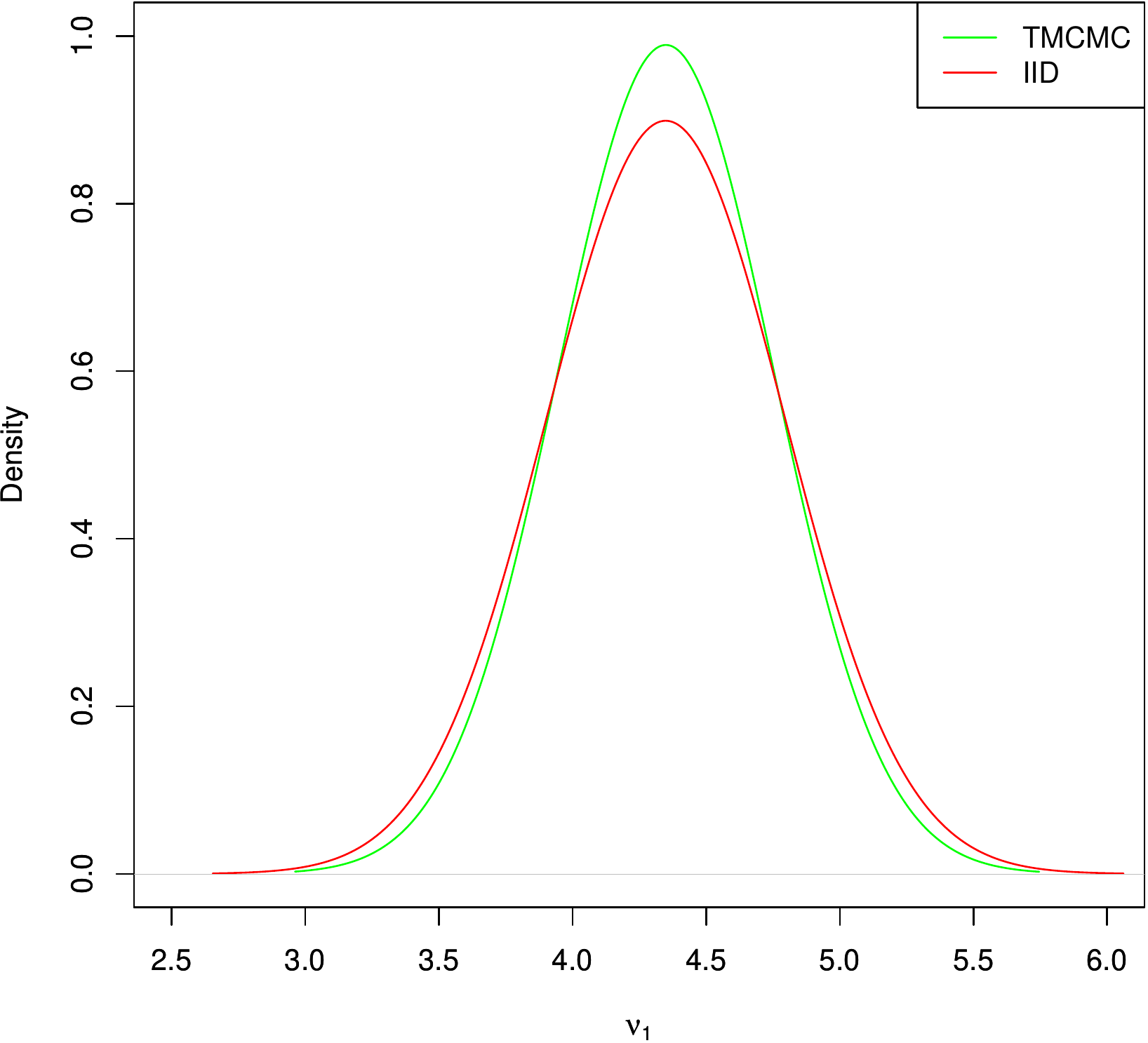}}
	\hspace{2mm}
	\subfigure []{ \label{fig:nu2}
	\includegraphics[width=6.0cm,height=6.0cm]{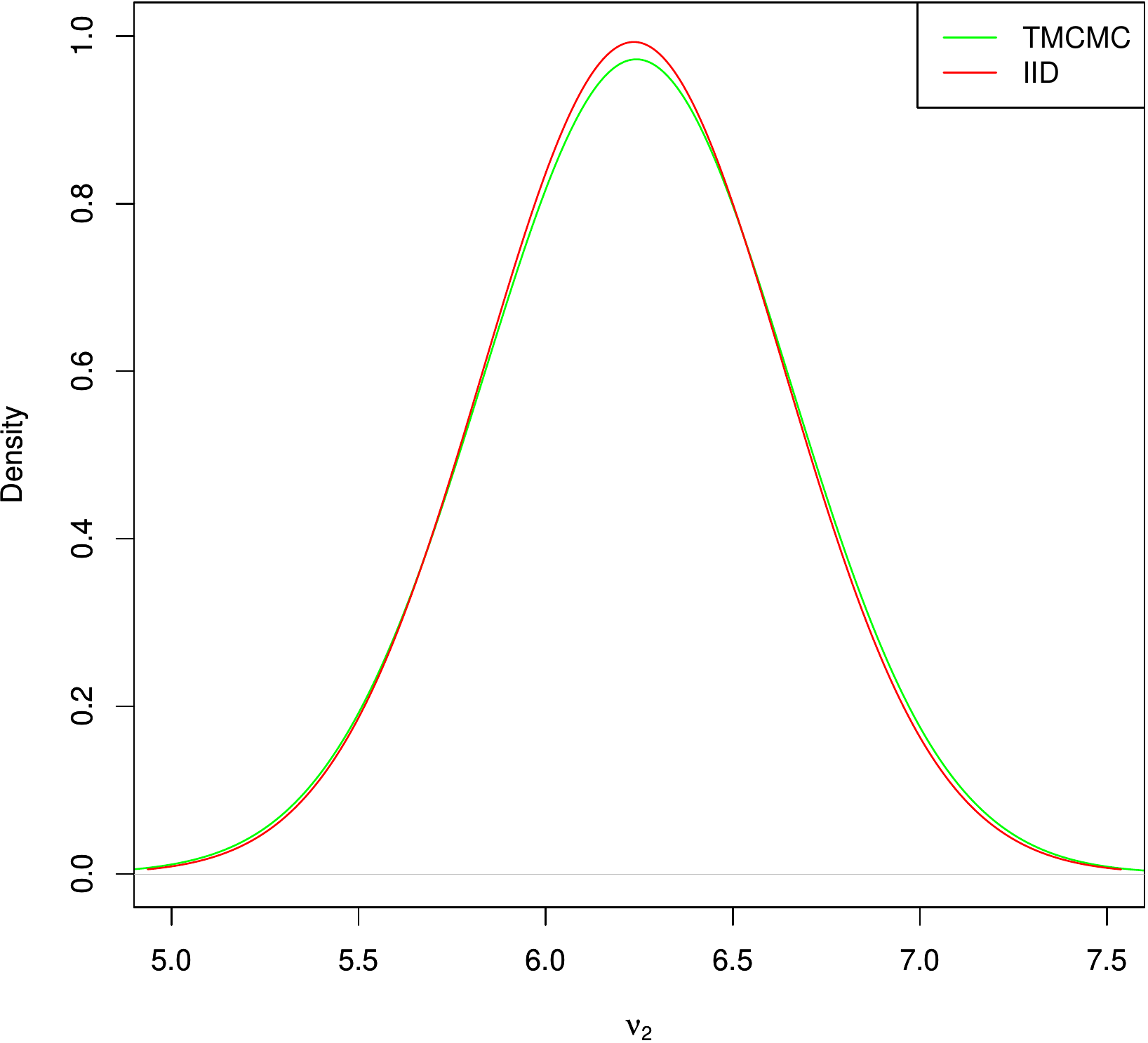}}\\
	\vspace{2mm}
	\subfigure []{ \label{fig:tau1}
	\includegraphics[width=6.0cm,height=6.0cm]{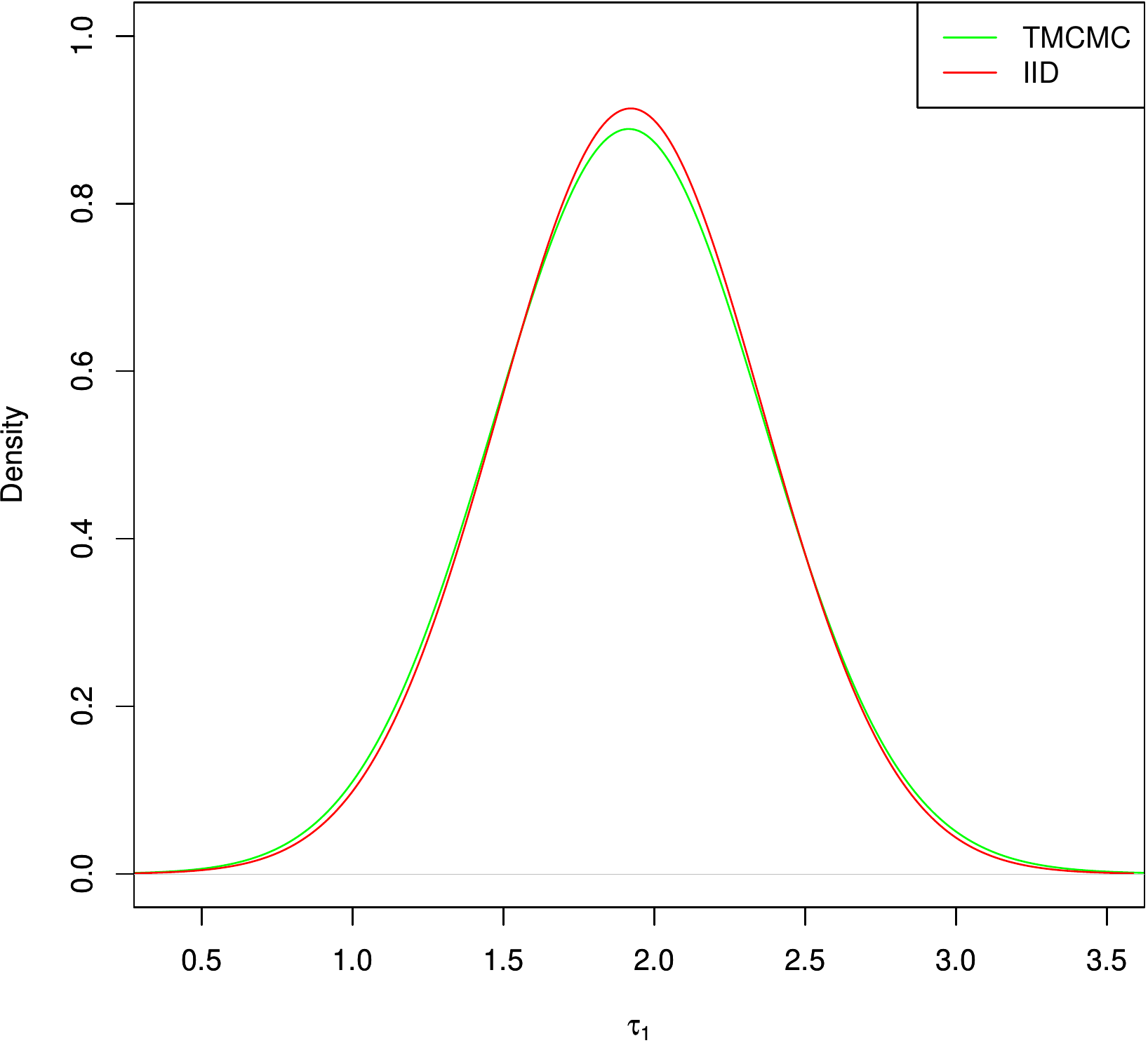}}
	\hspace{2mm}
	\subfigure []{ \label{fig:tau2}
	\includegraphics[width=6.0cm,height=6.0cm]{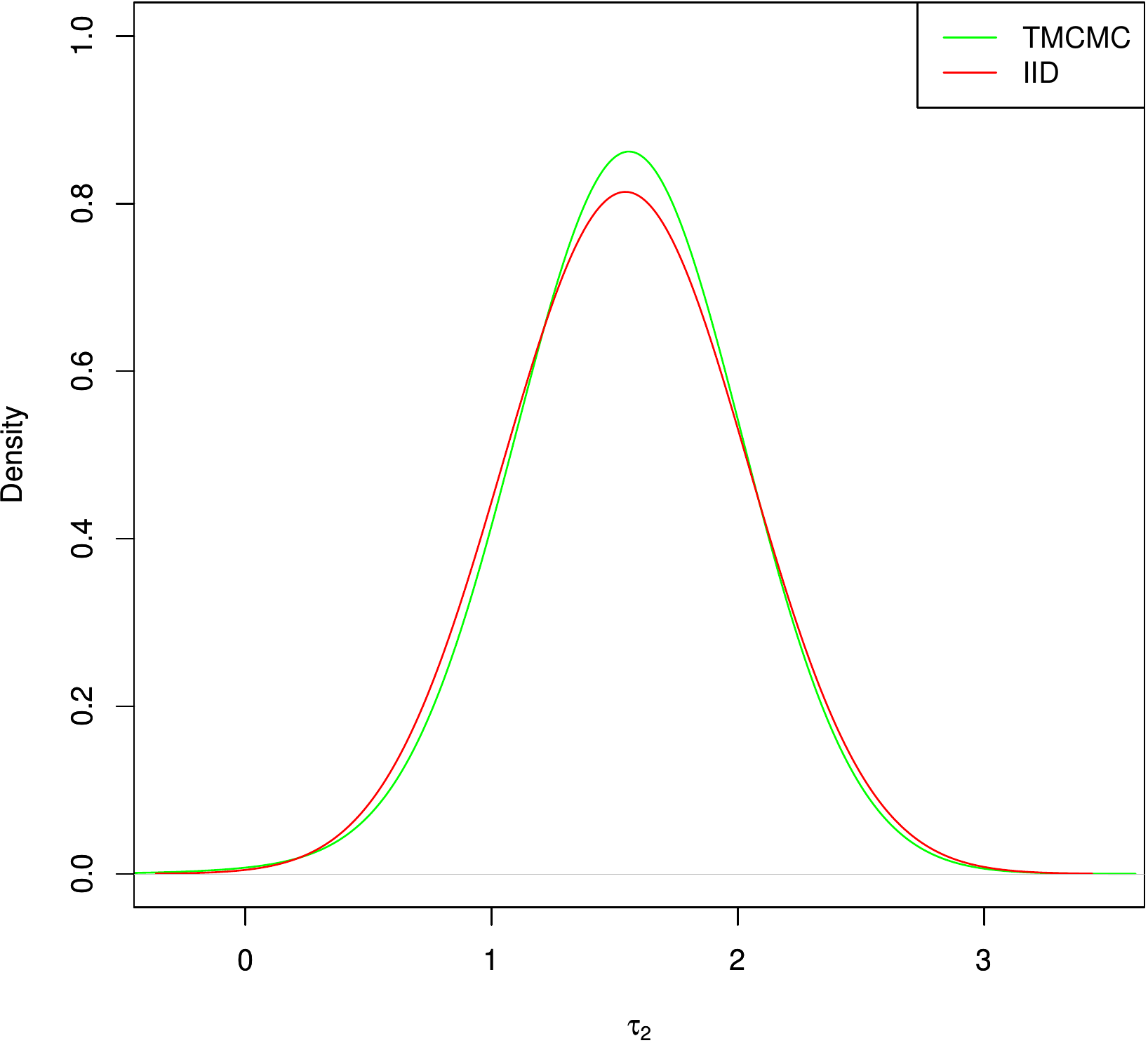}}\\
	\vspace{2mm}
	\subfigure []{ \label{fig:w1}
	\includegraphics[width=6.0cm,height=6.0cm]{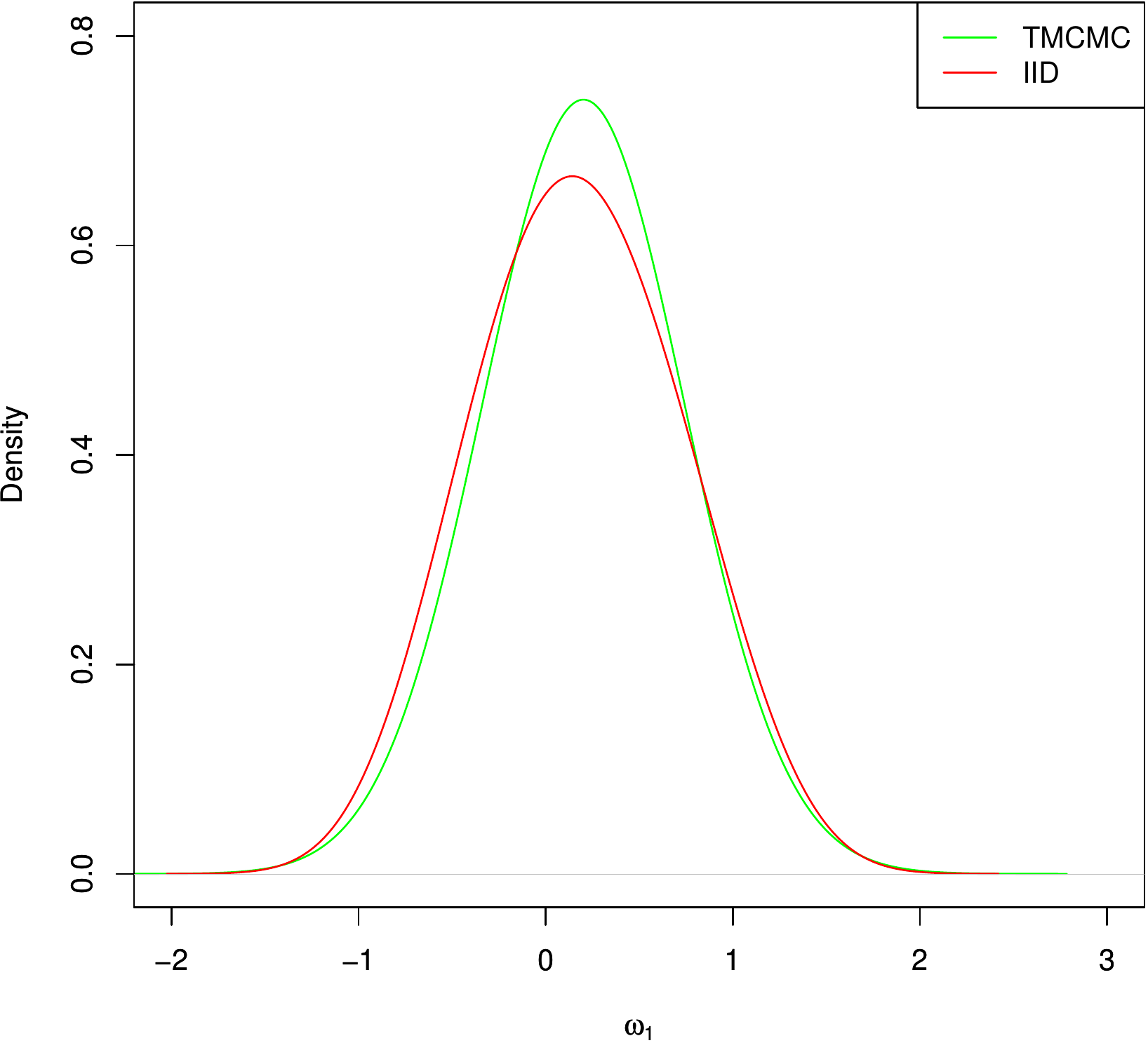}}
	\hspace{2mm}
	\subfigure []{ \label{fig:w2}
	\includegraphics[width=6.0cm,height=6.0cm]{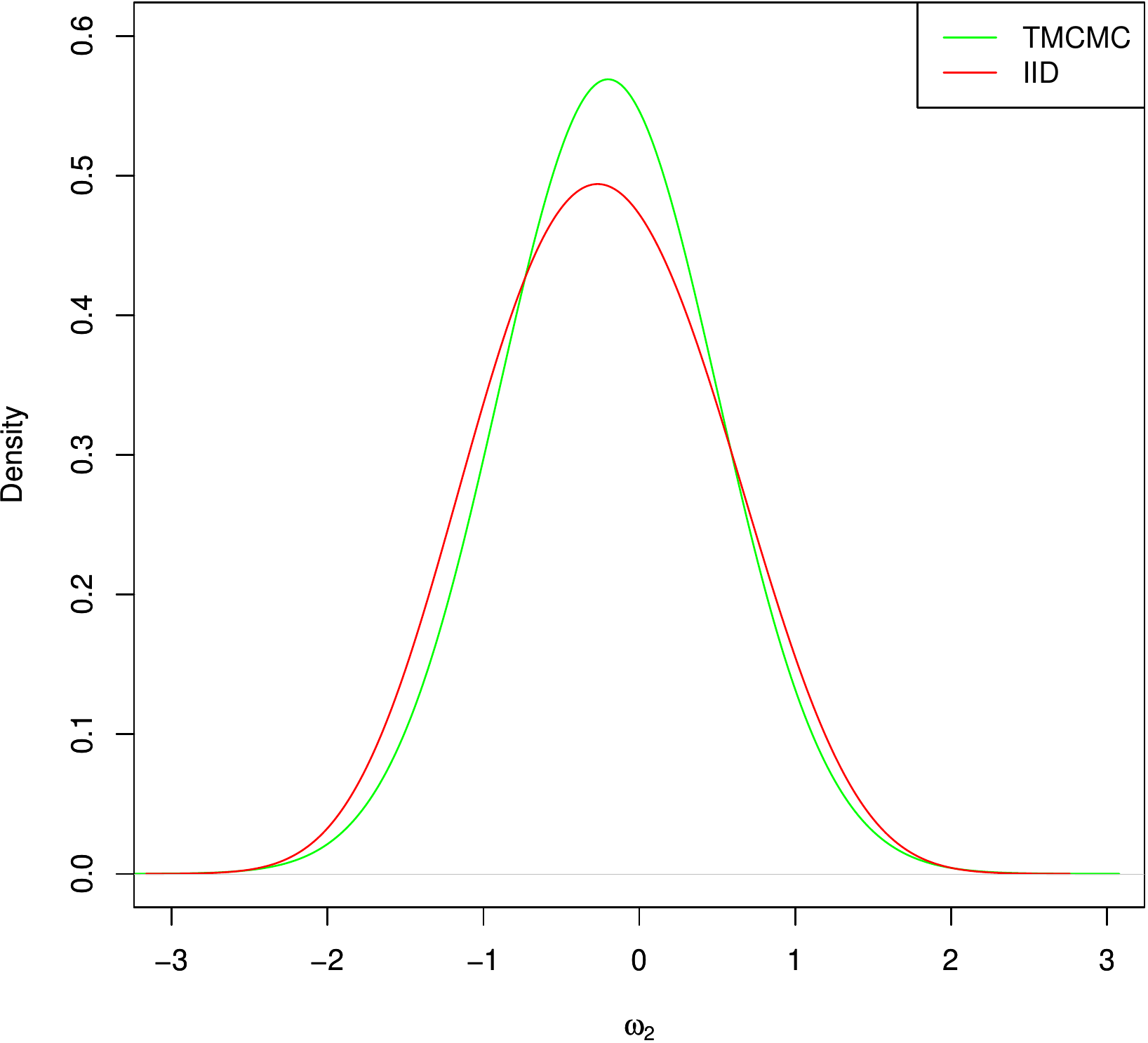}}
	\caption{TMCMC and $iid$ based posterior $\pi(\btheta_k|k=2,\by)$. 
	The red and green colours denote the $iid$ sample based density and TMCMC based density, respectively.}
	\label{fig:acidity_2comp}
\end{figure}

Figure \ref{fig:acidity_2comp} displays the marginal densities of the parameters corresponding to $\pi(\btheta_k|k=2,\by)$ based on TMCMC and $iid$ samplers.
It is to be seen that although TMCMC and $iid$ sampling are essentially in agreement, in general $iid$ sampling tends to explore the tail regions slightly
better than TMCMC. 

\begin{figure}
	\centering
	\subfigure []{ \label{fig:nu1_3}
	\includegraphics[width=4.8cm,height=4.8cm]{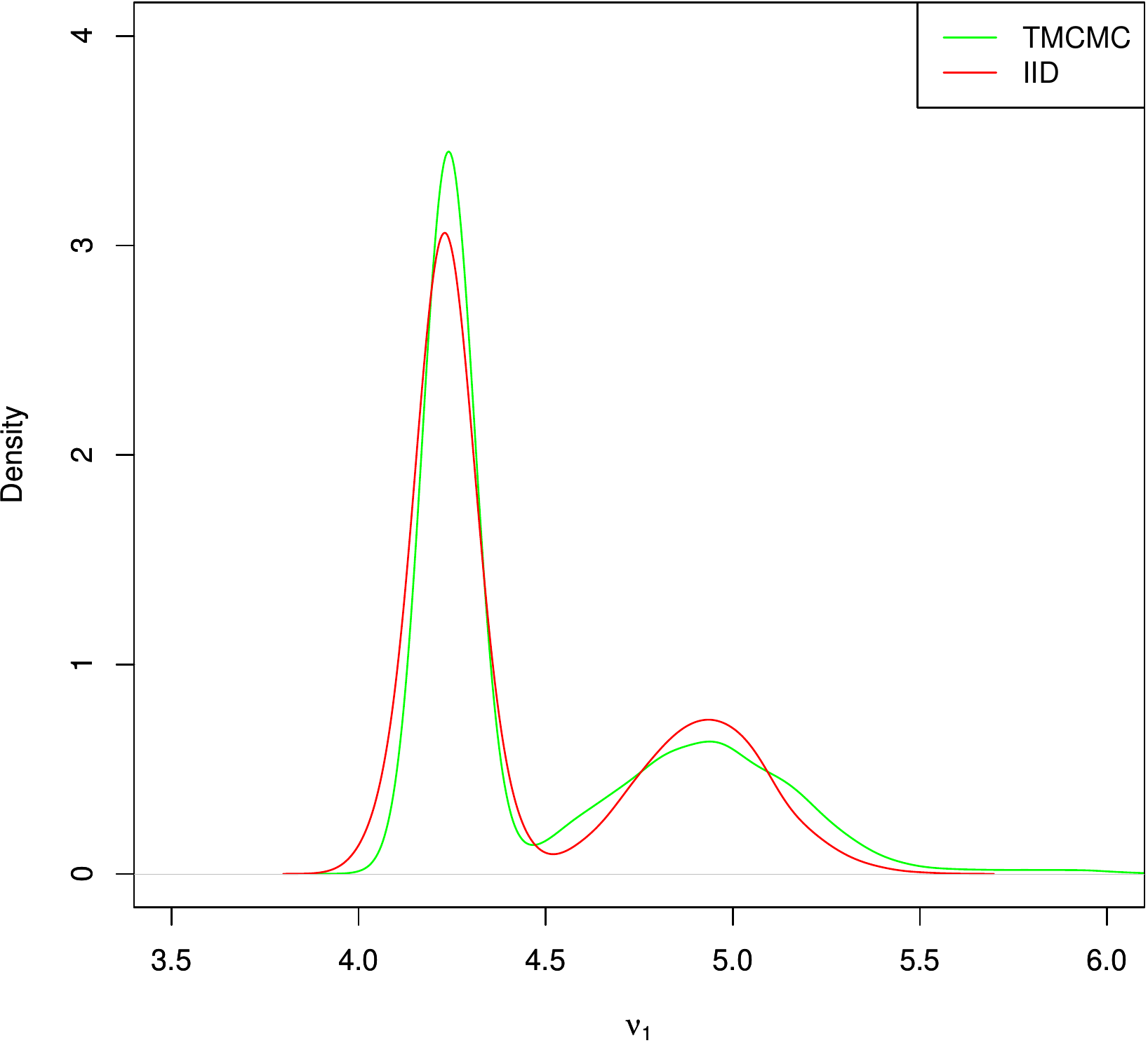}}
	\hspace{2mm}
	\subfigure []{ \label{fig:nu2_3}
	\includegraphics[width=4.8cm,height=4.8cm]{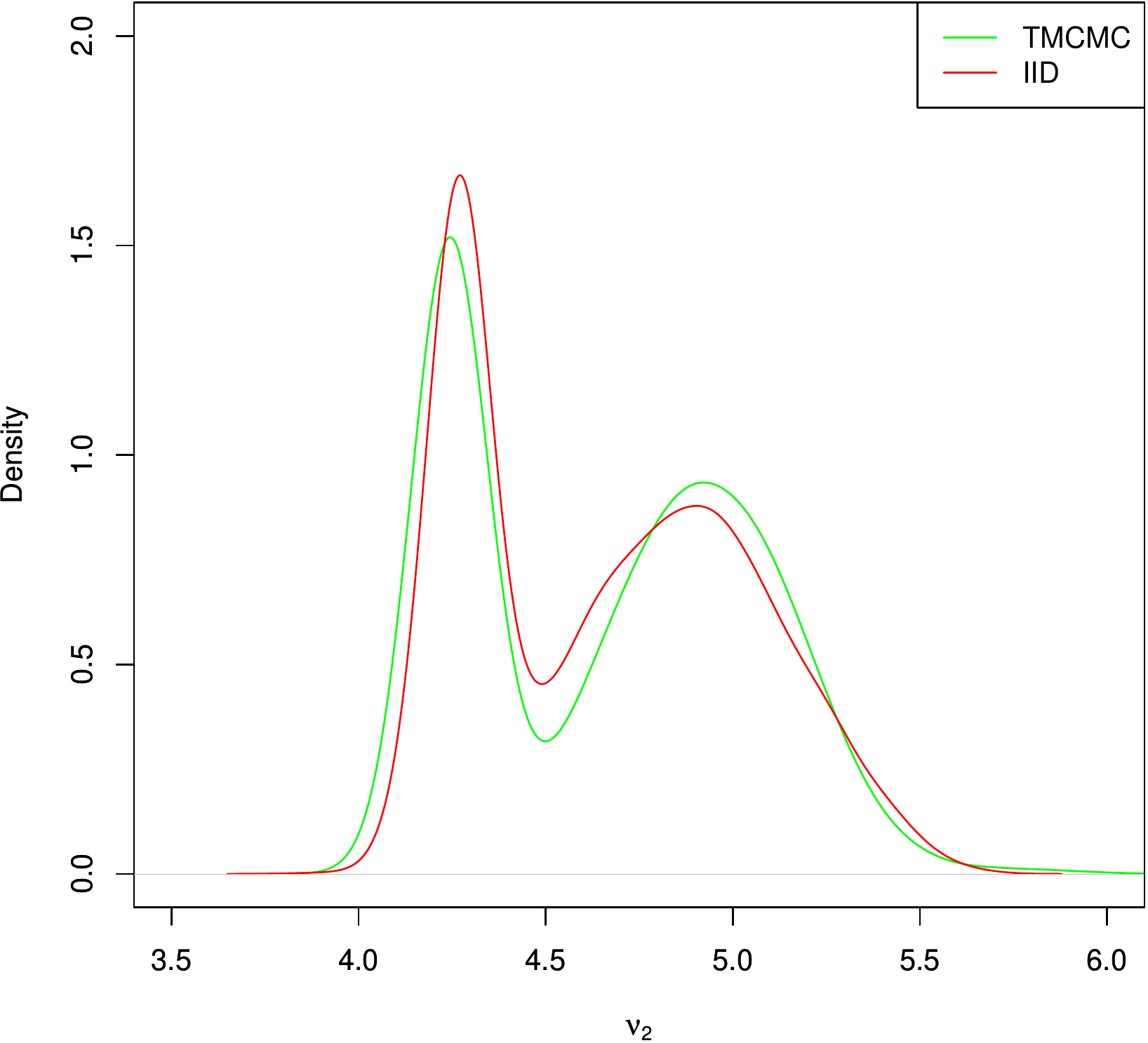}}
	\hspace{2mm}
	\subfigure []{ \label{fig:nu3_3}
	\includegraphics[width=4.8cm,height=4.8cm]{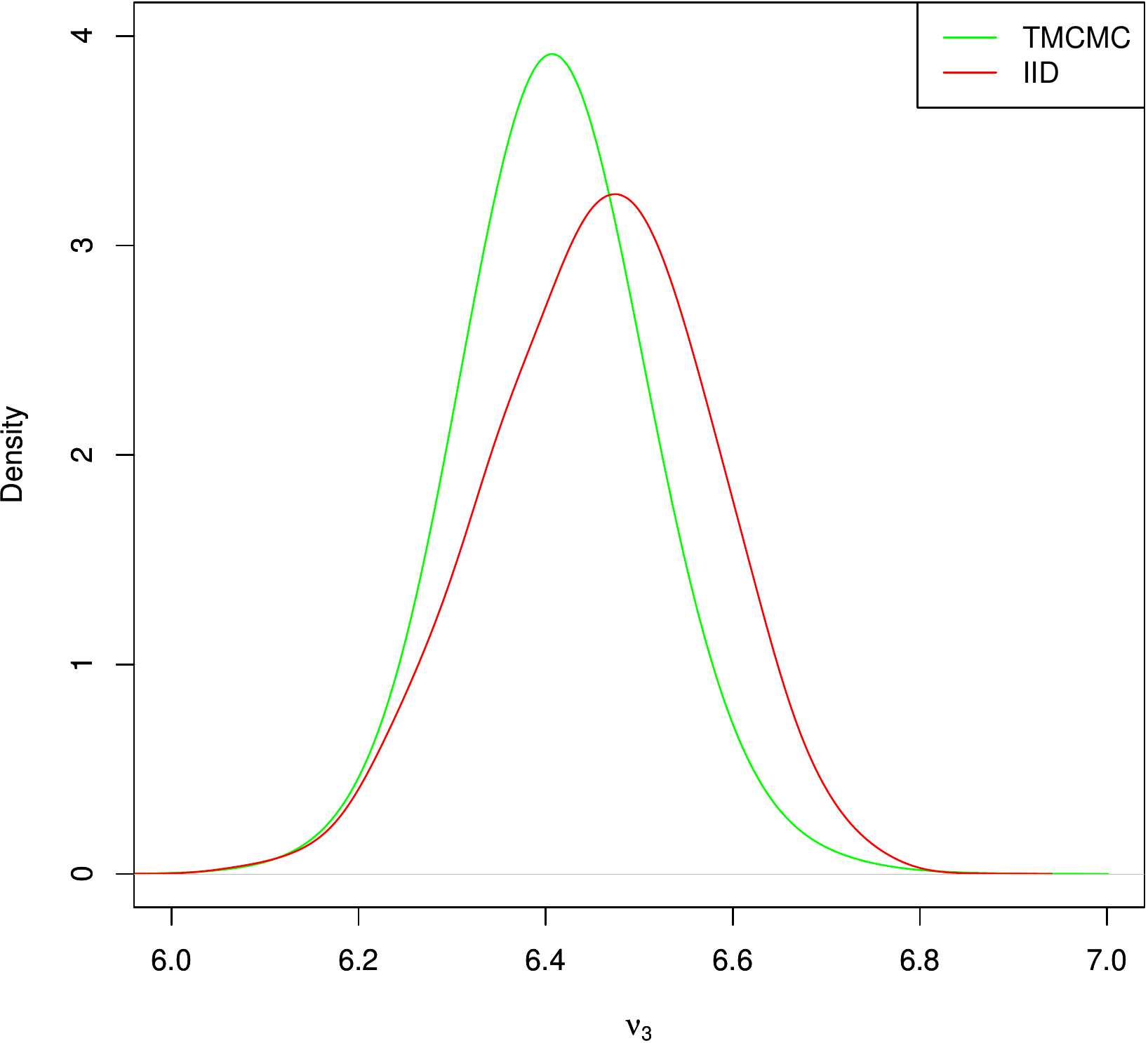}}\\
	\vspace{2mm}
	\subfigure []{ \label{fig:tau1_3}
	\includegraphics[width=4.8cm,height=4.8cm]{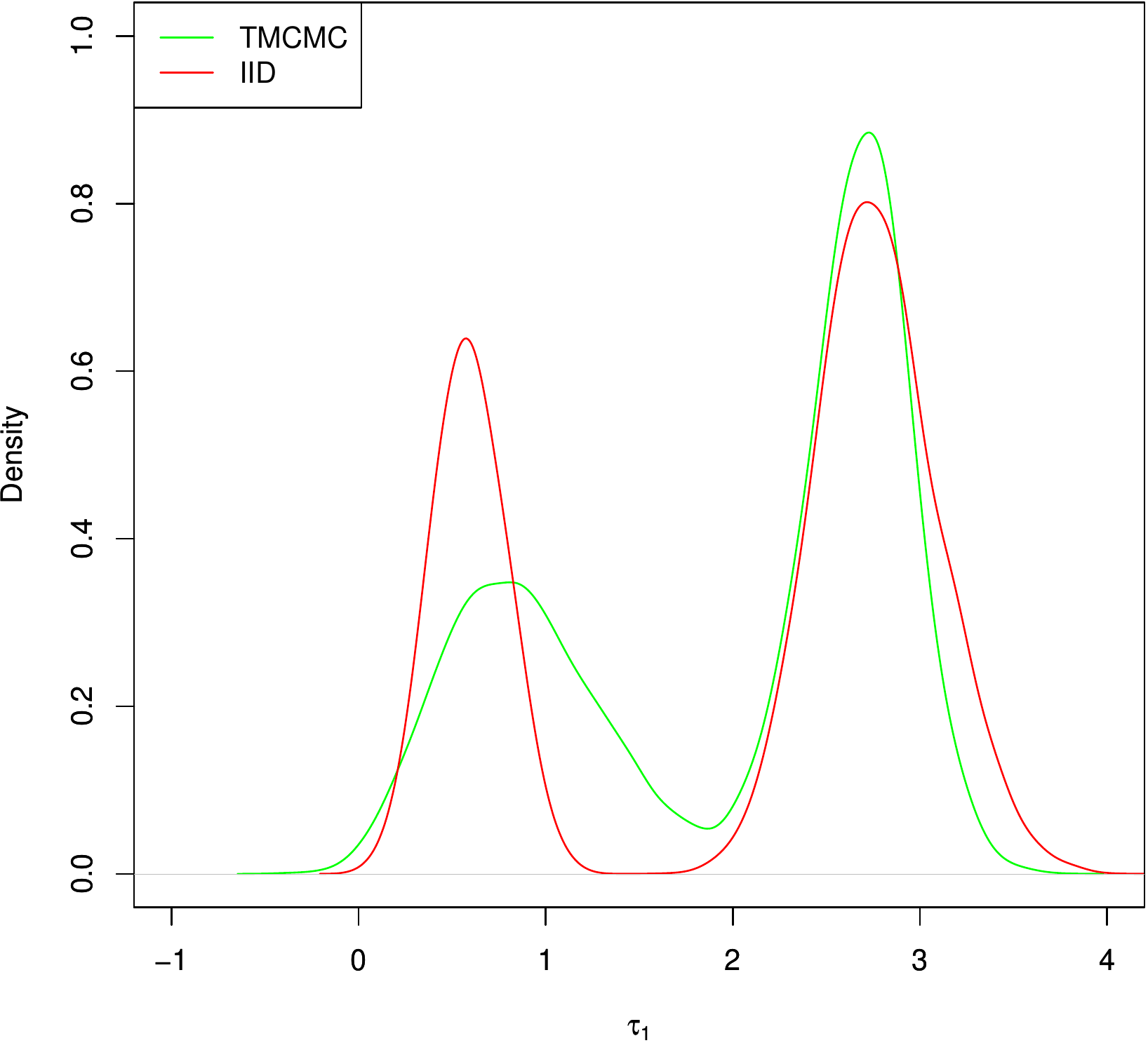}}
	\hspace{2mm}
	\subfigure []{ \label{fig:tau2_3}
	\includegraphics[width=4.8cm,height=4.8cm]{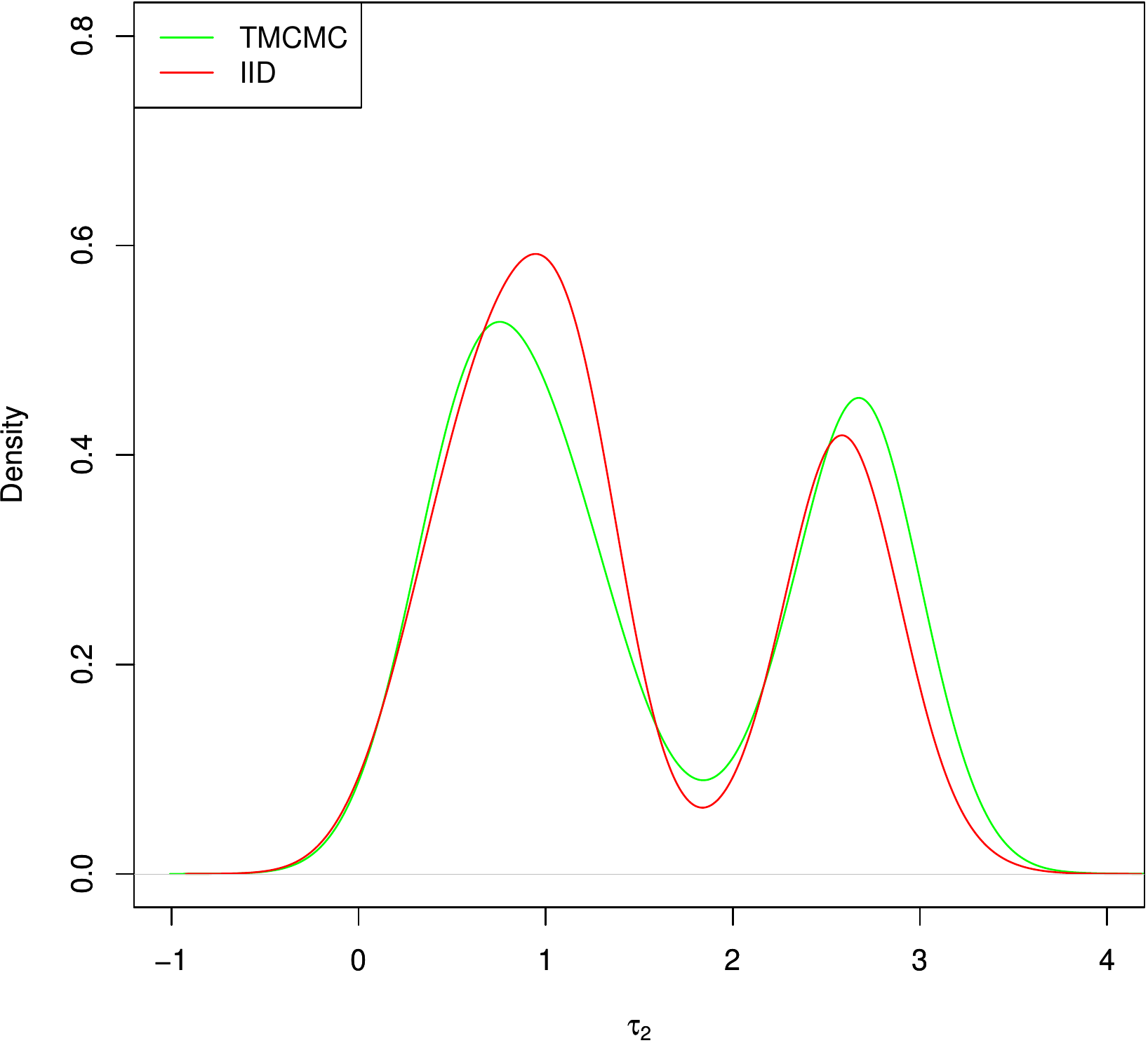}}
	\hspace{2mm}
	\subfigure []{ \label{fig:tau3_3}
	\includegraphics[width=4.8cm,height=4.8cm]{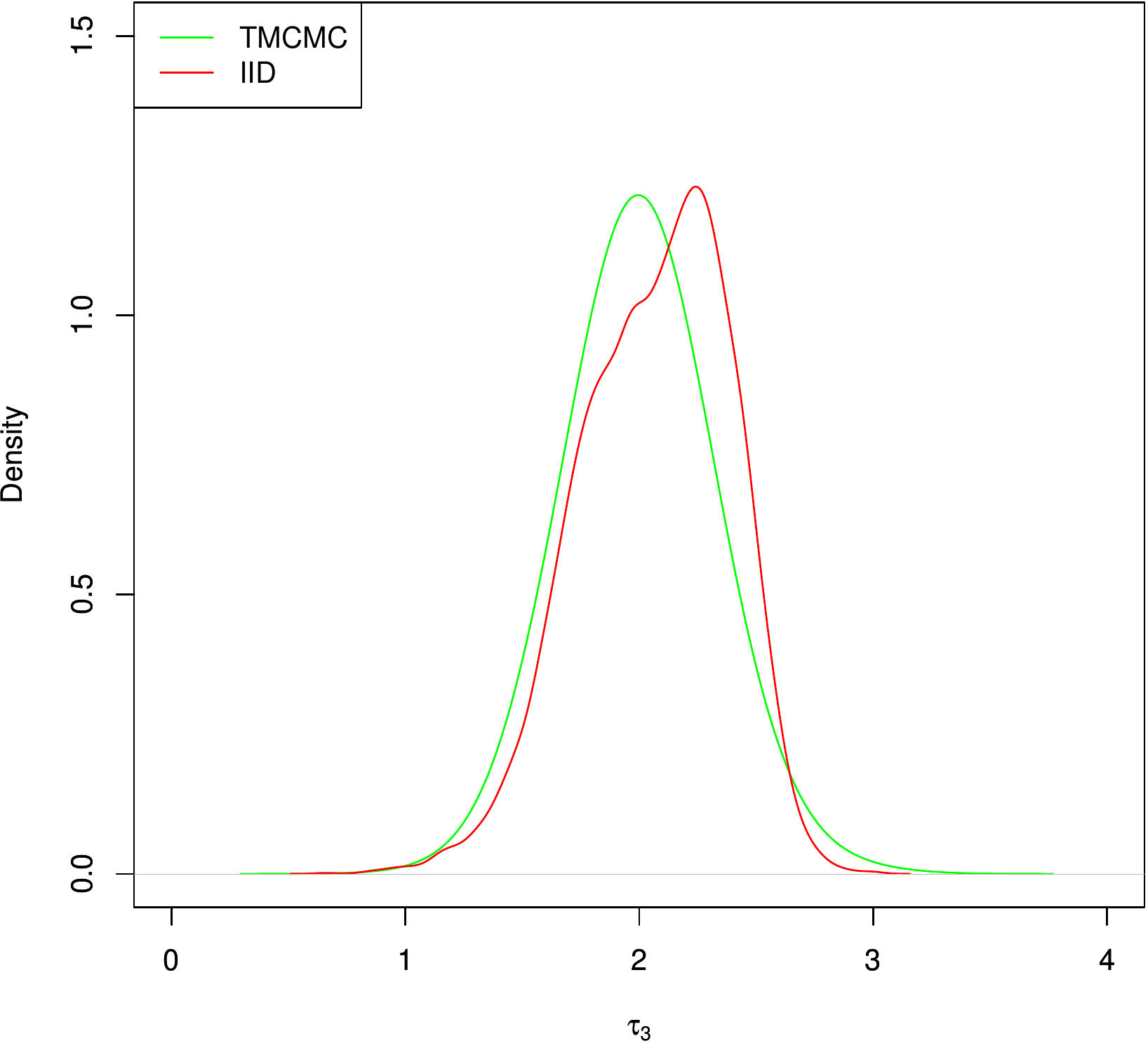}}\\
	\vspace{2mm}
	\subfigure []{ \label{fig:w1_3}
	\includegraphics[width=4.8cm,height=4.8cm]{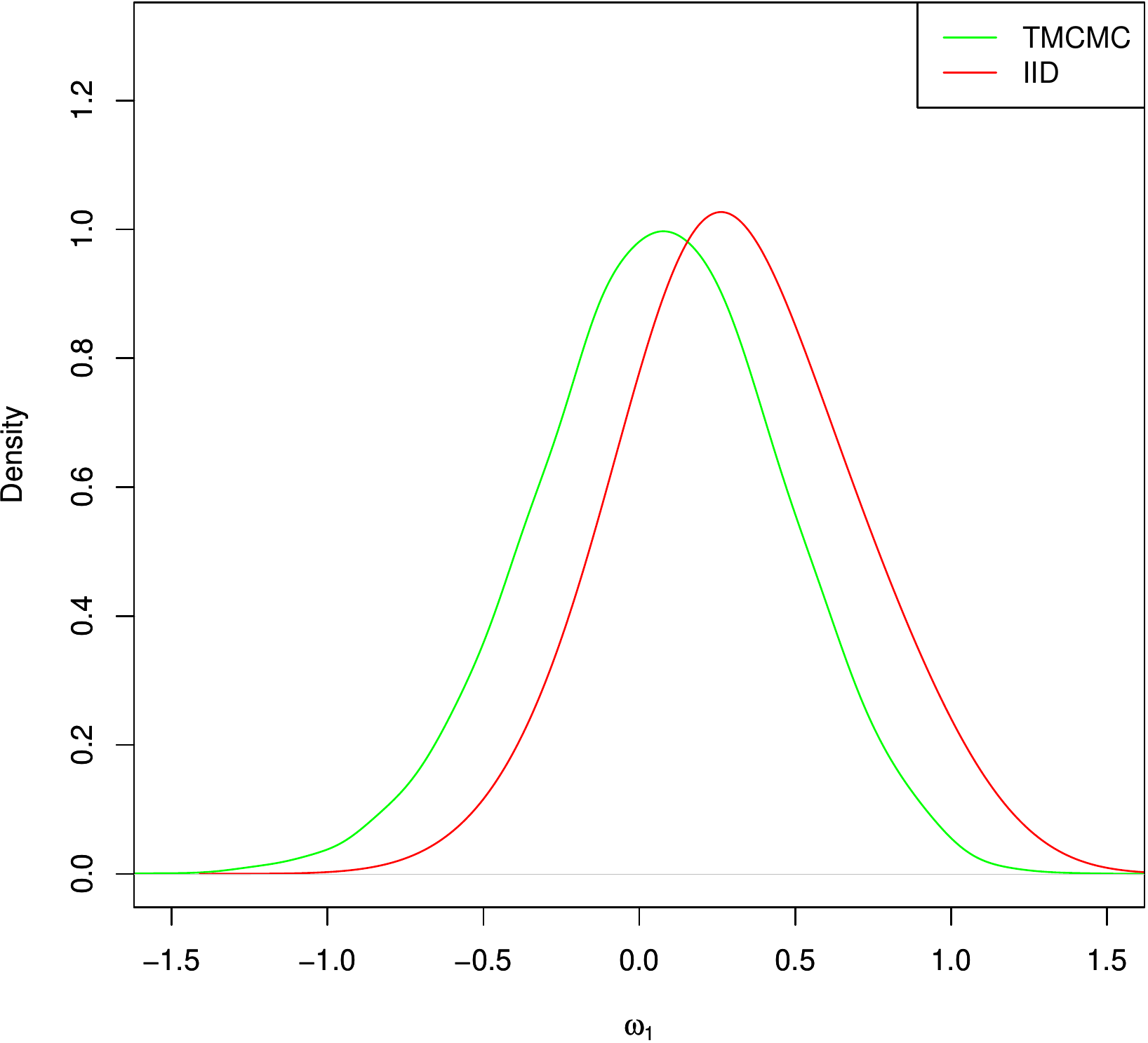}}
	\hspace{2mm}
	\subfigure []{ \label{fig:w2_3}
	\includegraphics[width=4.8cm,height=4.8cm]{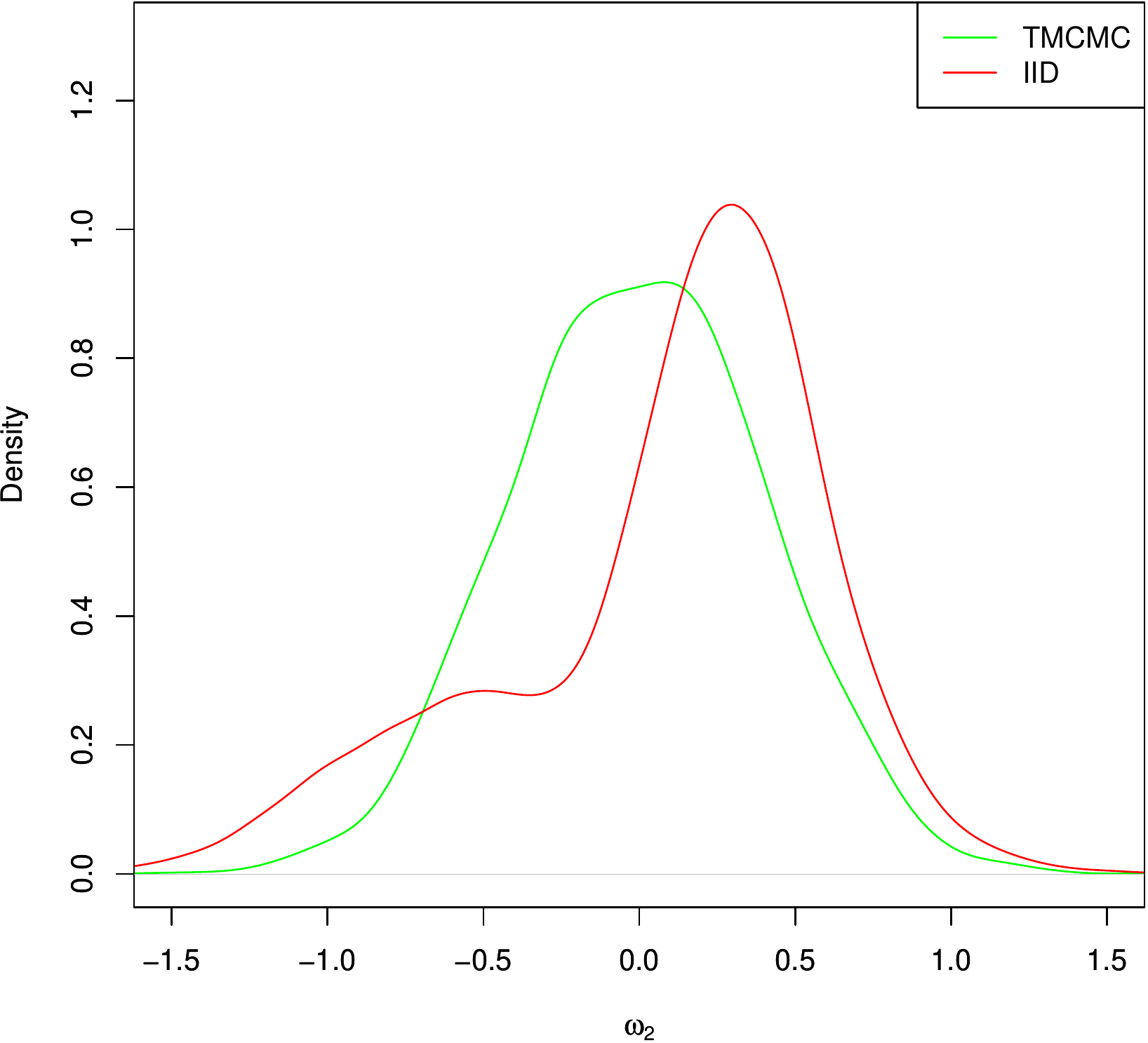}}
	\hspace{2mm}
	\subfigure []{ \label{fig:w3_3}
	\includegraphics[width=4.8cm,height=4.8cm]{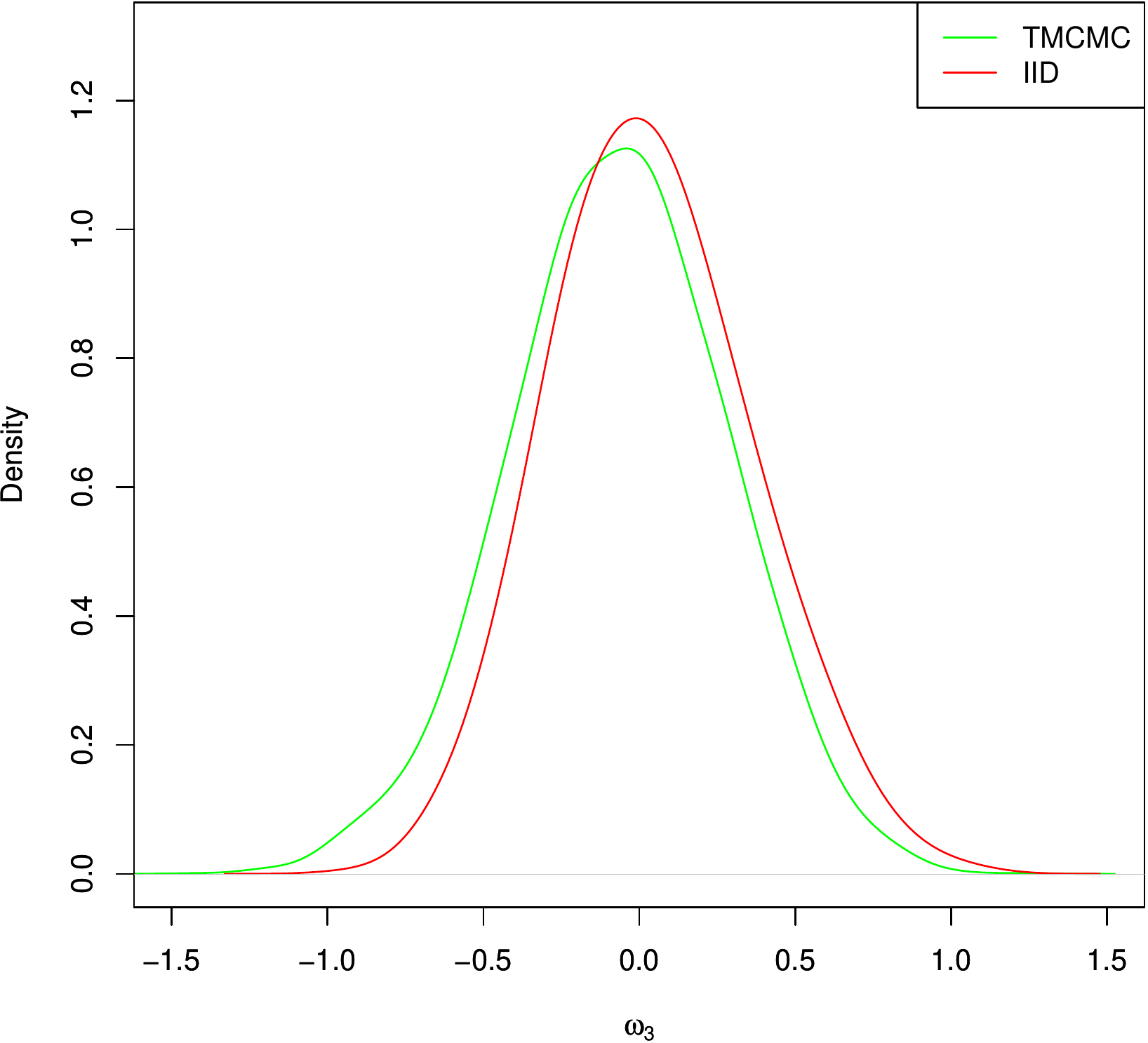}}
	\caption{TMCMC and $iid$ based posterior $\pi(\btheta_k|k=3,\by)$. 
	The red and green colours denote the $iid$ sample based density and TMCMC based density, respectively.}
	\label{fig:acidity_3comp}
\end{figure}

Figure \ref{fig:acidity_3comp} displays the TMCMC and $iid$ based marginal densities of the parameters corresponding to $\pi(\btheta_k|k=3,\by)$.
Observe that again TMCMC and $iid$ sampling are essentially in agreement, but the disagreements here are more stark than in the unimodal
case of $\pi(\btheta_k|k=2,\by)$. In particular, for $\omega_2$ (panel \ref{fig:w2_3}), TMCMC seems to have missed a minor left mode.
This might be responsible for the discrepancy between TMCMC and $iid$ cases for $\omega_1$ (panel \ref{fig:w1_3}) and $\omega_3$ (panel \ref{fig:w3_3}).
The discrepancy in panel \ref{fig:tau1_3} may be the result of TMCMC staying stuck at the left mode somewhat longer than desired, while that in panel \ref{fig:nu3_3}
is due to better exploration of the tail regions by the $iid$ sampler.
The differences in the other panels are relatively minor and may be attributed to the Markovian nature of TMCMC as opposed to the $iid$ sampler.

\begin{figure}
\includegraphics[width=6in,height=4.5in]{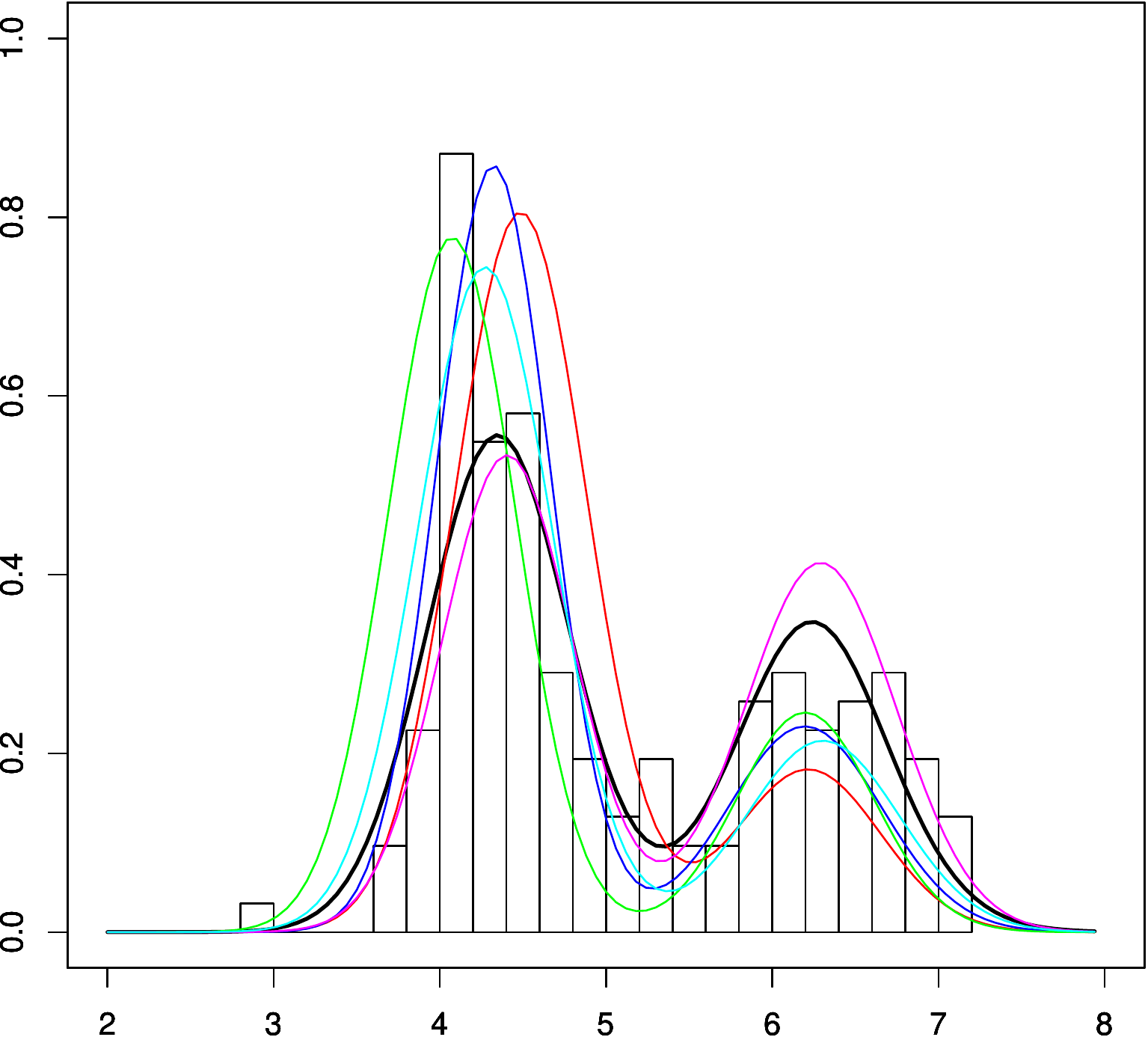}
\caption{Goodness of fit of the posterior distribution of densities (coloured curves) 
to the observed data (histogram). The thick black curve is the expected density of the posterior predictive distribution.}
\label{fig:acidity_hist}
\end{figure}

Figure \ref{fig:acidity_hist} shows the histogram of the acidity data and some density curves associated with the posterior predictive distribution, given by
\begin{equation}
\pi(\tilde y|\by)=\sum_{k\in\mathcal I}\int_{\mathbb R^{d_k}} f(\tilde y|\btheta_k,k)\pi(k|\by)\pi(\btheta_k|k,\by)d\btheta_k,
\label{eq:postpred}
\end{equation}
for $\tilde y\in\mathbb R$. In our case, for each equispaced $\tilde y=2+0.06\times i$, for $i=0,1,\ldots,99$ in the relevant range $[2,8]$, 
partitioned into $100$ sub-intervals, we obtain $\pi(\tilde y|\by)$ by plugging the $iid$ simulations from 
$\pi(k|\by)$ and $\pi(\btheta_k|k,\by)$ into $f(\tilde y|\btheta_k,k)$ given by (\ref{eq:normix}), which constitute the sample based posterior predictive distribution of the
densities. Some of these are overlapped on the histogram of Figure \ref{fig:acidity_hist}. The thick black curve is the point-by-point 
average of all the sample based densities, which is the Monte Carlo estimate of the expected density corresponding to (\ref{eq:postpred}). 
Satisfactory fit to the data is indicated by this diagram. The corresponding TTMCMC based figure can be found in \ctn{Das19}.

\section{Summary and conclusion}
\label{sec:conclusion}
\ctn{Bhatta21}, by creating a novel methodology for $iid$ sampling from arbitrary distributions of any dimension on the Euclidean space,
had paved the way for further development in the realm of multimodal distributions and variable-dimensional scenarios. Both these situations are 
considered extremely challenging, even for the purpose of MCMC sampling, letting alone $iid$ simulations. In this work, we have attempted to
extend the work of \ctn{Bhatta21} to accommodate these two setups, and show that $iid$ sampling is not only possible in such cases, but very much achievable,
in spite of the challenges that are usually considered insurmountable.

The key extension for multimodal distributions considered here is the identification of the modes of the underlying (high-dimensional) distribution, 
with the help of some existing works,
namely, those of \ctn{Sabya11}, \ctn{Das19} and \ctn{Roy20}, and then constructing the modal regions and the associated mixing probabilities. These information
are then integrated into Algorithm 1 of \ctn{Bhatta21} in a way that ensures validity and efficiency of the resultant $iid$ sampling procedure.
The crux of the idea is to choose the modal regions with the corresponding mixing probabilities, represent the target distribution as an infinite mixture
on the ellipsoid and annuli developed by the modal region, and then apply the perfect sampling procedure on the mixture sampled using the associated mixture
probabilities corresponding to the infinite representation.
It is important to appreciate that appropriate choice of the diffeomorphism parameter plays a crucial role with regard to efficiency of the perfect sampling strategy.
We also developed a diffeomorphism based Monte Carlo estimation procedure of the mixing probabilities of the infinite mixture representation, where again the
diffeomorphism parameter plays an important role.

With regard to the variable-dimensional setup, the key idea is to consider $iid$ sampling from $\pi(k|\by)$ and $\pi(\btheta_k|k,\by)$. Although the 
$iid$ sampling procedure for the latter is no different from that proposed in \ctn{Bhatta21} or Algorithm \ref{algo:perfect} for multimodal distributions,
sampling from $\pi(k|\by)$ requires first integrating out $\btheta_k$ from $f(\by|\btheta_k,k)\pi(\btheta_k|k)$. 
We showed how efficient Monte Carlo estimate of the resultant integral can be obtained by combining the Monte Carlo estimates on the relevant
ellipsoids and annuli, aided by appropriate diffeomorphisms. The entire procedure is shown to be highly amenable to parallel processing.

We illustrated our proposed theories and methods for multimodal setups with an example of a $50$-dimensional two-component normal mixture, obtaining quite encouraging results.
With a real, acidity data, we have also illustrated the versatility and efficacy of our $iid$ sampling procedure for variable-dimensional cases.

However, a somewhat disconcerting issue for variable dimensions is that, if a large number of values is supported by the posterior of $k$, then $iid$ sampling
is necessary for a large number of posteriors of the form $\pi(\btheta_k|k,\by)$. Since for different values of $k$, different tunings are necessary with respect
to choices of ellipsoids and annuli, diffeomorphism parameters, choices of appropriate modal regions for multimodal cases, it might require an enormous amount of manual
labour to generate $iid$ samples efficiently from such variable-dimensional distributions. In our future endeavor, we shall attempt to automate such tunings. 


\renewcommand\baselinestretch{1.3}
\normalsize
\bibliography{irmcmc}

\begin{thebibliography}{}

\bibitem[Bhattacharya(2008)Bhattacharya]{Bhattacharya08}
Bhattacharya, S. (2008).
\newblock {G}ibbs {S}ampling {B}ased {B}ayesian {A}nalysis of {M}ixtures with
  {U}nknown {N}umber of {C}omponents.
\newblock {\em Sankhya. Series B\/}, {\bf 70}, 133--155.

\bibitem[Bhattacharya(2021a)Bhattacharya]{Bhattacharya21}
Bhattacharya, S. (2021a).
\newblock {B}ayesian {L}\'{e}vy-{D}ynamic {S}patio-{T}emporal {P}rocess:
  {T}owards {B}ig {D}ata {A}nalysis.
\newblock arXiv:2105.08451v1.

\bibitem[Bhattacharya(2021b)Bhattacharya]{Bhatta21}
Bhattacharya, S. (2021b).
\newblock {IID} {S}ampling from {I}ntractable {D}istributions.
\newblock arXiv preprint.

\bibitem[Das and Bhattacharya(2019)Das and Bhattacharya]{Das19}
Das, M. and Bhattacharya, S. (2019).
\newblock {T}ransdimensional {T}ransformation {B}ased {M}arkov {C}hain {M}onte
  {C}arlo.
\newblock {\em Brazilian Journal of Probability and Statistics\/}, {\bf 33}(1),
  87--138.

\bibitem[Das and Bhattacharya(2020)Das and Bhattacharya]{Das20}
Das, M. and Bhattacharya, S. (2020).
\newblock {N}onstationary, {N}onparametric, {N}onseparable {B}ayesian
  {S}patio-{T}emporal {M}odeling {U}sing {K}ernel {C}onvolution of {O}rder
  {B}ased {D}ependent {D}irichlet {P}rocess.
\newblock arXiv:1405.4955v2.

\bibitem[Dutta and Bhattacharya(2014)Dutta and Bhattacharya]{Dutta14}
Dutta, S. and Bhattacharya, S. (2014).
\newblock {M}arkov {C}hain {M}onte {C}arlo {B}ased on {D}eterministic
  {T}ransformations.
\newblock {\em Statistical Methodology\/}, {\bf 16}, 100--116.
\newblock Also available at http://arxiv.org/abs/1106.5850. Supplement
  available at http://arxiv.org/abs/1306.6684.

\bibitem[Green(1995)Green]{Green95}
Green, P.~J. (1995).
\newblock {R}eversible jump {M}arkov chain {M}onte {C}arlo computation and
  {B}ayesian model determination.
\newblock {\em Biometrika\/}, {\bf 82}, 711--732.

\bibitem[Johnson and Geyer(2012)Johnson and Geyer]{Johnson12a}
Johnson, L.~T. and Geyer (2012).
\newblock {V}ariable {T}ransformation to {O}btain {G}eometric {E}rgodicity in
  the {R}andom-{W}alk {M}etropolis {A}lgorithm.
\newblock {\em The Annals of Statistics\/}, {\bf 40}, 3050--3076.

\bibitem[Mukhopadhyay and Bhattacharya(2021)Mukhopadhyay and
  Bhattacharya]{Minerva21}
Mukhopadhyay, M. and Bhattacharya, S. (2021).
\newblock {B}ayes {F}actor {A}symptotics for {V}ariable {S}election in the
  {G}aussian {P}rocess {F}ramework.
\newblock {\em Annals of the Institute of Statistical Mathematics\/}.
\newblock To appear.

\bibitem[Mukhopadhyay {\em et~al.}(2011)Mukhopadhyay, Bhattacharya, and
  Dihidar]{Sabya11}
Mukhopadhyay, S., Bhattacharya, S., and Dihidar, K. (2011).
\newblock {O}n {B}ayesian ``{C}entral {C}lustering'': {A}pplication to
  {L}andscape {C}lassification of {W}estern {G}hats.
\newblock {\em Annals of Applied Statistics\/}, {\bf 5}, 1948--1977.

\bibitem[Propp and Wilson(1996)Propp and Wilson]{Prop96}
Propp, J.~G. and Wilson, D.~B. (1996).
\newblock {E}xact {S}ampling with {C}oupled {M}arkov {C}hains and
  {A}pplications to {S}tatistical {M}echanics.
\newblock {\em Random Structures and Algorithms\/}, {\bf 9}, 223--252.

\bibitem[Richardson and Green(1997)Richardson and Green]{Richardson97}
Richardson, S. and Green, P.~J. (1997).
\newblock {O}n {B}ayesian {A}nalysis of {M}ixtures with an {U}nknown {N}umber
  of {C}omponents (with discussion).
\newblock {\em Journal of the Royal Statistical Society. Series B\/}, {\bf 59},
  731--792.

\bibitem[Roy and Bhattacharya(2020)Roy and Bhattacharya]{Roy20}
Roy, S. and Bhattacharya, S. (2020).
\newblock {F}unction {O}ptimization with {P}osterior {G}aussian {D}erivative
  {P}rocess.
\newblock arXiv:2010.13591v1.

\end{thebibliography}


\end{document}